\documentclass[aps,pre,preprint]{revtex4-2}
\usepackage{amsmath}
\usepackage{graphicx}
\usepackage{bbm}
\usepackage{subfigure}
\usepackage[caption=false]{subfig}
\begin{document}
\title{Plasmonic Dielectric Response of Finite Temperature Electron Gas}
\author{M. Akbari-Moghanjoughi}
\affiliation{Faculty of Sciences, Department of Physics, Azarbaijan Shahid Madani University, 51745-406 Tabriz, Iran\\ massoud2002@yahoo.com}

\begin{abstract}
In this research we report the dielectric response of a finite temperature electron gas, electrostatically interacting with both external and self-induced plasmonic fields, in the well-known random phase approximation. The generalized energy dispersion relation which incorporates the plasmonic band structure is used to calculate the Lindhard dielectric response of homogenous electron gas from which many important physical functionals, such as the structure factor, loss function, screening potential, optical reflectivity and electronic stopping power are deduced. The present dual length-scale theory of dielectric response incorporates both single electron as well as collective electrostatic oscillation of electrons which, due to the Van-Hove-like singularity at plasmon wavenumber, shows distinct features of plasmonic response to electromagnetic interactions of the electron gas with arbitrary degree of electron degeneracy. It is shown that the static impurity charge screening potential is oscillatory Lennard-Jones-type attractive potential which is quite different from both predicted by the Conventional noninteracting Lindhard theory and the Shukla-Eliasson attractive potential obtained from quantum hydrodynamic approach. It is also revealed that due to resonant electron-plasmon interactions and multi-pole structure of the electronic response function the Landau damping region is scattered. The findings of current research may have important implications for a wide range of physical phenomena relevant to a broad nonrelativistic electron density-temperature regime, from the laboratory scale semiconductors and nanoelectronic technology to the warm dense matter state.
\end{abstract}

\date{\today}

\maketitle
\newpage

\section{Introduction}

The study of light- and particle-matter interactions are of primary subjects of many interdisciplinary physical and chemical sciences \cite{lucio}. Such studies have been the cornerstones of fundamental modern physics theories development during the past century such as quantum electronics, optoelectronics and plasmonics. Fundamental theories which our technological endeavors like miniaturized semiconductor technology and emerging fields of opto- and nano-electronics \cite{mark,haug}, plasmonics \cite{gardner,man1,maier}, low dimensional quantum devices \cite{yofee}, etc. strongly rely on. In plasmas external perturbation techniques are the main tools in stability diagnostics and provide fundamental probes to measure the thermodynamic state quantities such as temperature, density, collision rates and scattering parameters \cite{chen,krall}. The linear response theory \cite{ichimaru1,ichimaru2,ichimaru3} is one of the most effective approximate methods to investigate dynamic behavior of a wide range of many body statistical systems subject to external perturbations in the laboratory scale \cite{fetter,mahan,pin}. Linear approximations are useful due to lack of exact analytic models in treatment of large quantity interacting quantum statistical systems. However, with the invent of fast computing devices and improved algorithms many computational platforms based on quantum theories such as density functional \cite{axel}, Hartree-Fock procedure \cite{fischer}, quantum hydrodynamic models \cite{haas1,man2} and quantum kinetic simulation methods \cite{man3} has been developed and refined over the recent years in order to probe features of interacting quantum statistical systems beyond the conventional response theories.

Quantum effects give rise to fascinating unique physical properties of metallic compounds and semiconductors \cite{kit,ash,hu1,seeg} which are appealing in semiconductor and nanotechnologies. They also play fundamental role in many physical processes in warm dense matter (WDM) \cite{ko} and astrophysical dense environements \cite{chandra}. Pioneering developments \cite{madelung,fermi,hoyle,bohm,bohm1,bohm2,pines,levine,klimontovich} of quantum statistical theories over past few decades have led to emergence of modern quantum plasma theories. Recent studies of linear and nonlinear aspects of dense plasmas based on quantum kinetic, hydrodynamic and magnetohydrodynamic theories \cite{se,sten,ses,brod1,mark1,man4,fhaas,scripta,stenf1,stenf2,stenf3,stenf4}, density functional oriented hydrodynamic models \cite{bonitz} and Schr\"{o}dinger-Poisson Madelung fluid system \cite{manfredi,hurst}, have revealed interesting new aspects of quantum plasmas which are not present in classical counterparts. The new findings also confirm that these theoretical tools provide quite useful means of studying quantum statistical systems with large degree of freedom and various charged species.

It is however well-known that in order to capture the essence of kinetic aspects of the statistical systems such as the Landau damping effect and nonlinear kinetic instabilities, one has to use the full kinetic approach such as the Vlaso-Poisson model. A rigorous kinetic treatment of quantum dielectric response of free electron gas has been developed by Lindhard within the framework of the random phase approximation (RPA) \cite{lind,stern}. The Lindhard dielectric response function is the most fundamental quantity in solid state theory from which many important physical quantities like dynamic and static structure factors \cite{sturm}, radial distribution function, dynamic and static charge screening potential, optical and electric dynamic conductivities, elastic and inelastic scattering cross-sections and stopping power, etc. are calculated \cite{ionst}. Within the RPA the complexity of many body electronic interactions is eliminated by considering the interaction of each independent electron with a self-consistent field. However, in the Lindhard theory of dielectric response electrons are considered as independent entities regarding their parabolic single electron energy dispersion. This is a very simplifying assumption which can lead to deviations from real experimental data because of ignorance of electrostatic interactions among electrons which can fundamentally modify the energy dispersion of electrons in collective excitations like plasmons. More recently, a dual length-scale theory of collective electron excitations \cite{akbquant, akbnew1, akbnew2} has been developed in which the plasmons carry two characteristic momentum, one due to the single-electron oscillations and the other due to their collective electrostatic interaction in the electron gas. It has been shown that the heat capacity of a quantum electron gas is greatly modified due to plasmonic effects at high electron gas temperatures \cite{akbheat}. Therefore, the generalized energy dispersion of electrons contain information not only due to single electron quantum states but also due the whole system interacting with the electrons via the internal electrostatic field. The later is equivalent to inclusion of effective mass due ro the plasmonic energy band structure in the conventional response theories which in the standard Lindhard dielectric response may be taken into account by using the generalized energy dispersion of electrons. In current improvement we use the Lindhard response theory within the framework of RPA by considering the generalized energy dispersion relation which accounts for the energy dependent effective electron mass in the plasmonic excitations.

\section{Linear Response of Homogenous Electron Gas}

When a homogenous electron gas with a neutralizing positive charge background is subject to an external field a time-dependent spatially inhomogenous charge density is $n_{ext}$ introduced. Then the electric displacement vector is given as
\begin{equation}\label{ed}
\nabla  \cdot {\cal{D}}(r,t) = {\rho_{ext}(r,t)}.
\end{equation}
The total charge density is then the combination of the external and the electrostatically induced charge densities
\begin{equation}\label{rho}
{\rho(r,t)} = {\rho_{ext}(r,t)}+{\rho_{ind}(r,t)},
\end{equation}
which relate to the corresponding electric field $\epsilon_0\nabla  \cdot {\cal{E}}(r,t) = -{\rho(r,t)}$ and scalar potentials ${\cal{E}}(r,t)=-\nabla\phi(r,t)$ and ${\cal{D}}(r,t)=-\epsilon_0\nabla\phi_{ext}(r,t)$.
The Fourier analysis of above equations, knowing that ${{\cal{D}}_\alpha }({\bf{q}},\omega ) = \sum\limits_\beta  {{\varepsilon _{\alpha \beta }}({\bf{q}},\omega )} {{\cal{E}}_\beta }({\bf{q}},\omega )$ for homogenous system, give rise to the following relationships
\begin{equation}\label{rel}
{\phi _{ext}}({\bf{q}},\omega ) = \bar \epsilon ({\bf{q}},\omega )\phi ({\bf{q}},\omega),\hspace{3mm}{n_{ext}}({\bf{q}},\omega ) = \bar \epsilon ({\bf{q}},\omega)\rho({\bf{q}},\omega ),
\end{equation}
where $\bar \epsilon ({\bf{q}},\omega )=\epsilon ({\bf{q}},\omega )/\epsilon_0$. Furthermore, the Fourier analysis of Poisson's equations lead to
\begin{equation}\label{pois}
\frac{1}{{\bar \varepsilon ({\bf{q}},\omega )}} = 1 + \frac{{4\pi {e^2}}}{{{q^2}}}\Pi ({\bf{q}},\omega),\hspace{3mm}\Pi ({\bf{q}},\omega ) = \frac{{{\rho_{ind}}({\bf{q}},\omega )}}{{{V_{ext}}({\bf{q}},\omega )}},
\end{equation}
where $V_{ext}=-e\phi_{ext}$ and $\Pi ({\bf{q}},\omega )$ is called the linear response function of the electron gas. The dielectric function in terms of the screened potential reads
\begin{equation}\label{sp}
\bar \varepsilon ({\bf{q}},\omega ) = 1 - \frac{{4\pi {e^2}}}{{{q^2}}}\bar \Pi ({\bf{q}},\omega ),\hspace{3mm}\bar \Pi ({\bf{q}},\omega ) = \frac{{{\rho _{ind}}({\bf{q}},\omega )}}{{{V_{sc}}({\bf{q}},\omega )}},
\end{equation}
where $V_{sc}=V_{ext}/\bar\epsilon$. On the other hand, the external potential can be viewed as a perturbation with corresponding Hamiltonian
\begin{equation}\label{ph}
{\cal{H}}_1 = \int {{V_{ext}}} ({\bf{r}},t)n({\bf{r}},t)d{\bf{r}}.
\end{equation}
The response function is also related to the density-density correlations via the Fourier transform
\begin{equation}\label{dd}
\Pi ({\bf{q}},\omega ) =  - \frac{i}{\hbar }\int\limits_0^\infty  {d(t - t')} {e^{i\omega (t - t') - \delta (t - t')}}\frac{1}{V}\left\langle {\left[ {n({\bf{q}},t),n( - {\bf{q}},t)} \right]} \right\rangle,
\end{equation}
where $\delta$ is a positive infinitesimal which ensures an adiabatic switching.

\section{Response Function in Random-Phase Approximation}

The thermally averaged linear response function is expressed as
\begin{equation}\label{rpr}
\Pi ({\bf{r}},{{\bf{r}}^\prime },\omega ) = \mathop {\lim }\limits_{\delta  \to {0^ + }} \sum\limits_{{\bf{mn}}} {\frac{{{f_{\bf{m}}} - {f_{\bf{n}}}}}{{\hbar (\omega  - {\omega _{{\bf{mn}}}}) + i\delta }}\Psi _{\bf{m}}^*({\bf{r}})\Psi _{\bf{n}}^*({{\bf{r}}^\prime }){\Psi _{\bf{m}}}({\bf{r}}){\Psi _{\bf{n}}}({{\bf{r}}^\prime })}.
\end{equation}
where $\Psi_{\bf{m}}({\bf{r}})$ is the many-body eigenstate at given position ${\bf{r}}$ and $\omega_{\bf{mn}}=\omega_{\bf{m}}-\omega_{\bf{n}}$ is the quasiparticle transition frequency jump. In Eq. (\ref{rpr}) $f_{\bf{m}}$ denotes the occupation probability of the quasiparticle excitation in the eigenstate {\bf{m}}. In the momentum space (\ref{rpr}) this translates into \cite{mih}
\begin{equation}\label{rpm}
\Pi ({\bf{q}},{{\bf{q}}^\prime },\omega ) = \mathop {\lim }\limits_{\delta  \to {0^ + }} \sum\limits_{{\bf{mn}}} {\frac{{{f_{\bf{m}}} - {f_{\bf{n}}}}}{{\hbar (\omega  - {\omega _{{\bf{mn}}}}) + i\delta }}\left\langle {{\bf{m}}\left| {{e^{i{\bf{q}}.{{\bf{r}}^\prime }}}} \right|{\bf{n}}} \right\rangle } \left\langle {{\bf{m}}\left| {{e^{i{{\bf{q}}^\prime }.{{\bf{r}}^{\prime \prime }}}}} \right|{\bf{n}}} \right\rangle,
\end{equation}
in which
\begin{equation}\label{dm}
\left\langle {{\bf{m}}\left| {{e^{i{\bf{q}}.{\bf{r}}}}} \right|{\bf{n}}} \right\rangle  = \int {\Psi _{\bf{m}}^*({\bf{r}}){e^{i{\bf{q}}.{\bf{r}}}}} {\Psi _{\bf{n}}}({\bf{r}}){\bf{dr}},
\end{equation}
are the density matrix elements. Therefore, the longitudinal electrostatic dielectric function in RPA is written as
\begin{equation}\label{df}
\epsilon ({\bf{q}},\omega ) = \epsilon_{\infty} - \frac{4\pi e^2}{q^2}\Pi ({\bf{q}},\omega ),
\end{equation}
where $\epsilon_{\infty}$ is the ambient dielectric constant. The Lindhard dielectric response function of the noninteracting \emph{free electron} gas within the RPA model is then given as \cite{lind}
\begin{equation}\label{ld}
\Pi ({\bf{q}},\omega ) = \frac{1}{{4{\pi ^2}}}\mathop {\lim }\limits_{\delta  \to {0^ + }} \int {\frac{{f({\bf{k}} + {\bf{q}}) - f({\bf{k}})}}{{\hbar (\omega  + i\delta ) - \left[\epsilon({\bf{k}} + {\bf{q}}) - \epsilon({\bf{k}})\right]}}} {\bf{dk}},
\end{equation}
in which $f({\bf{k}})=1/\{\exp[(\epsilon-\mu)/k_B T]+1\}$ is the Fermi-Dirac occupation function and $\epsilon({\bf{k}})=\hbar^2{\bf{k}}^2/2m^*$ is the single electron energy dispersion relation with $m^*$ being the effective electron mass which accounts for the band structure of semiconductors. The Lindhard dielectric function has been generalized by Mermin to include the electron-ion collision effects as
\begin{equation}\label{md}
{\epsilon _M}(q,\omega ) = 1 + \frac{{(\omega  + i\nu )\left[ {{\epsilon _{L}}({\bf{q}},\omega  + i\nu ) - 1} \right]}}{{\omega  + i\nu \left[ {{\epsilon _{L}}({\bf{q}},\omega  + i\nu ) - 1} \right]/\left[ {{\epsilon _{L}}({\bf{q}},0) - 1} \right]}},
\end{equation}
in which $\epsilon_L$ denotes the Lindhard dielectric function and $\nu$ is the electron-ion collision frequency \cite{soylom}. The conventional noninteracting \emph{free electron} response theory uses the concept of effective electron mass in free electron energy dispersion relation as a band structure correction to the model. However, such a modification can not go beyond the independent electron approximation and a new theory which accounts to the electron-electron interactions is needed to be developed in order to account for collective response of the electron gas. In the next section we obtain a generalized energy dispersion relation which accounts for the electrostatic interactions between the electrons in the system.

\section{Energy Dispersion Relation of Collective Excitations}

\begin{figure}[ptb]\label{Figure1}
\includegraphics[scale=0.6]{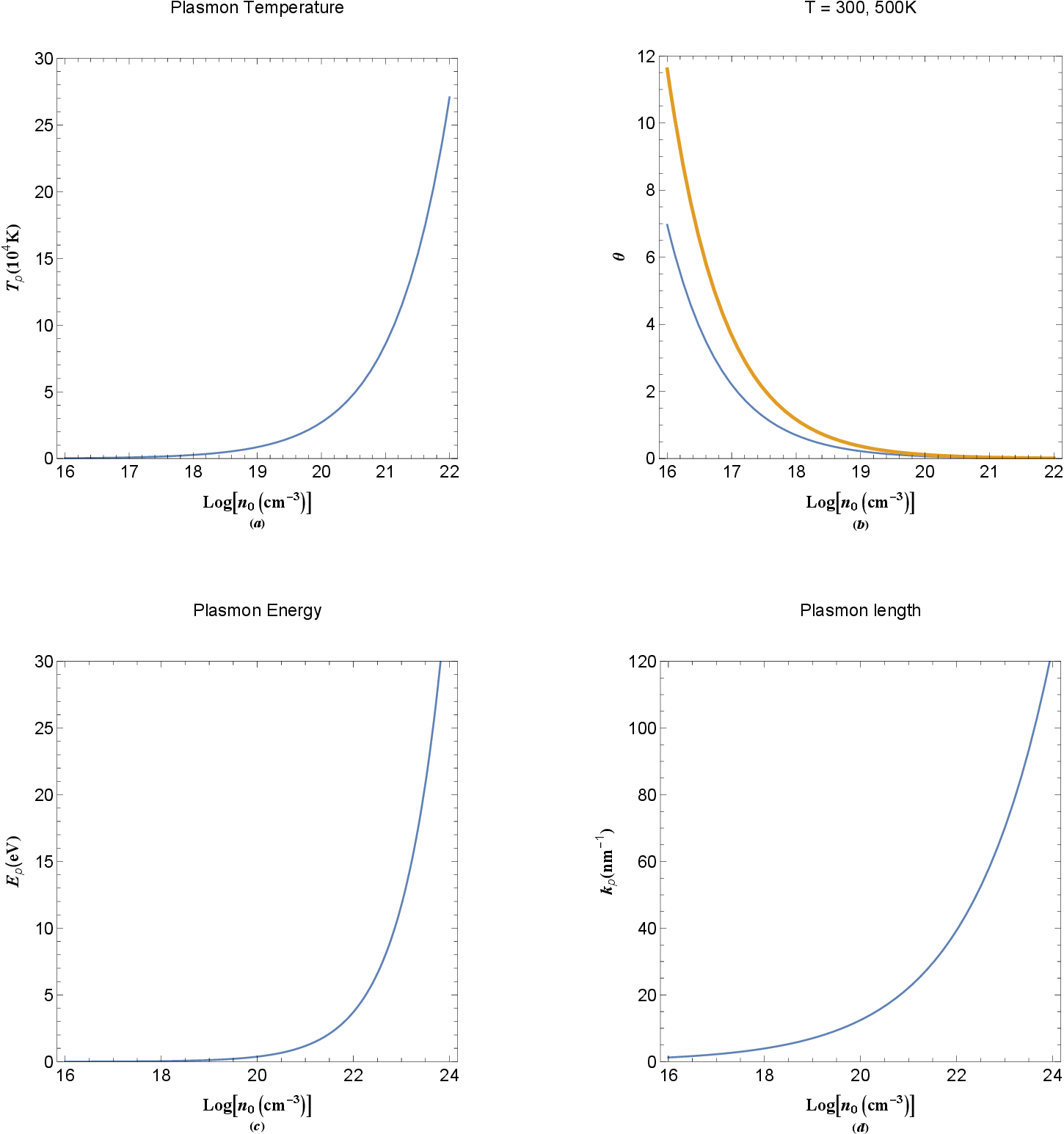}\caption{(a) The variation plasmon temperature with electron number-density. (b) The variation normalized electron temperature with electron number-density. (c) The variation plasmon energy with electron number-density. (d) The variation plasmon length with electron number-density. The thicker cureve in plot (b) refers to higher value above the panel.}
\end{figure}

\begin{figure}[ptb]\label{Figure2}
\includegraphics[scale=0.6]{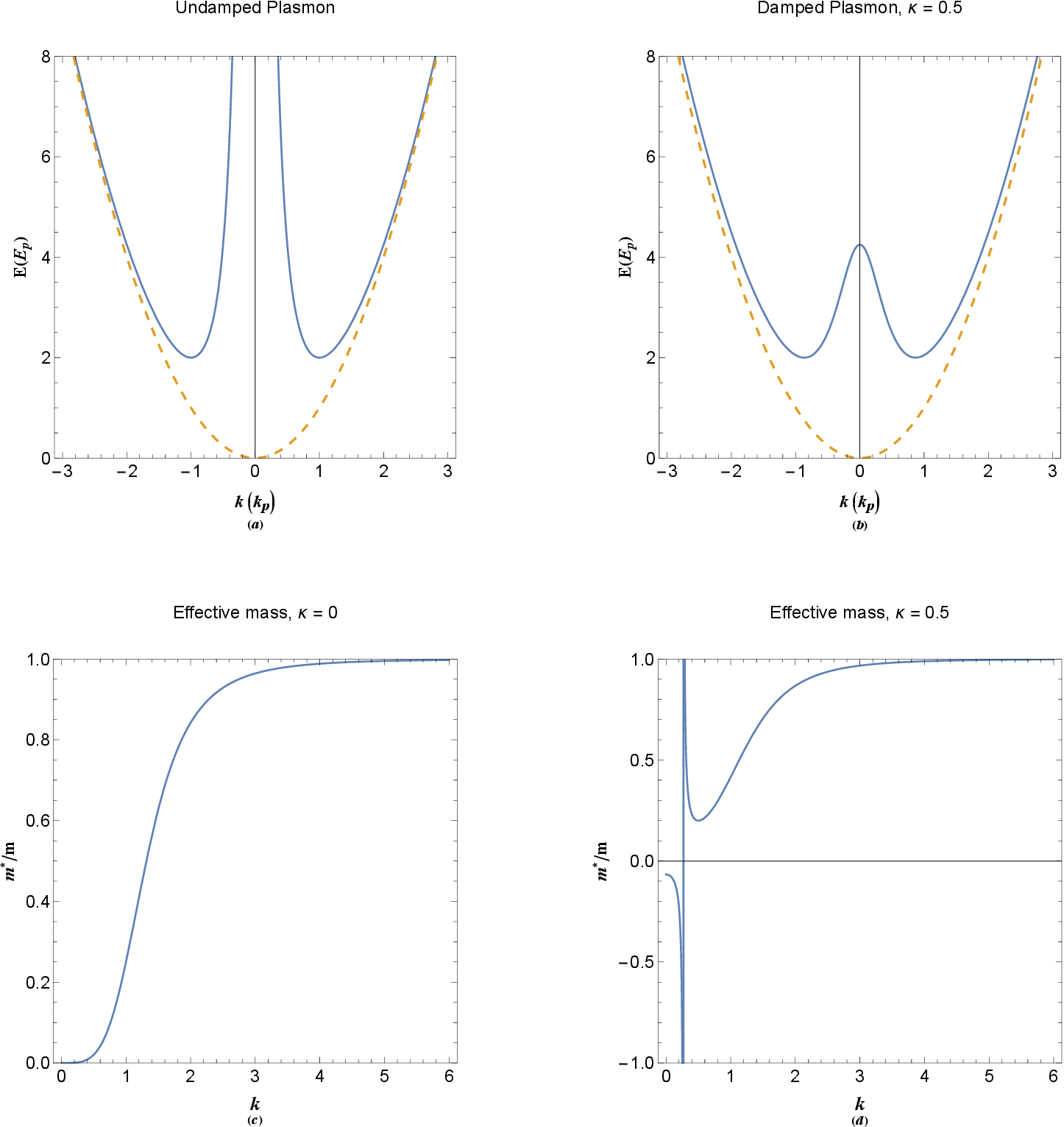}\caption{(a) The energy dispersion of undamped plasmon (solid curve) and free electron gas (dashed curve). (b) The energy dispersion of damped plasmon (solid curve) and free electron gas (dashed curve). (c) The effective electron mass of undamped plasmon excitations. (d) The effective mass of electron in damped plasmon excitations.}
\end{figure}

\begin{figure}[ptb]\label{Figure3}
\includegraphics[scale=0.56]{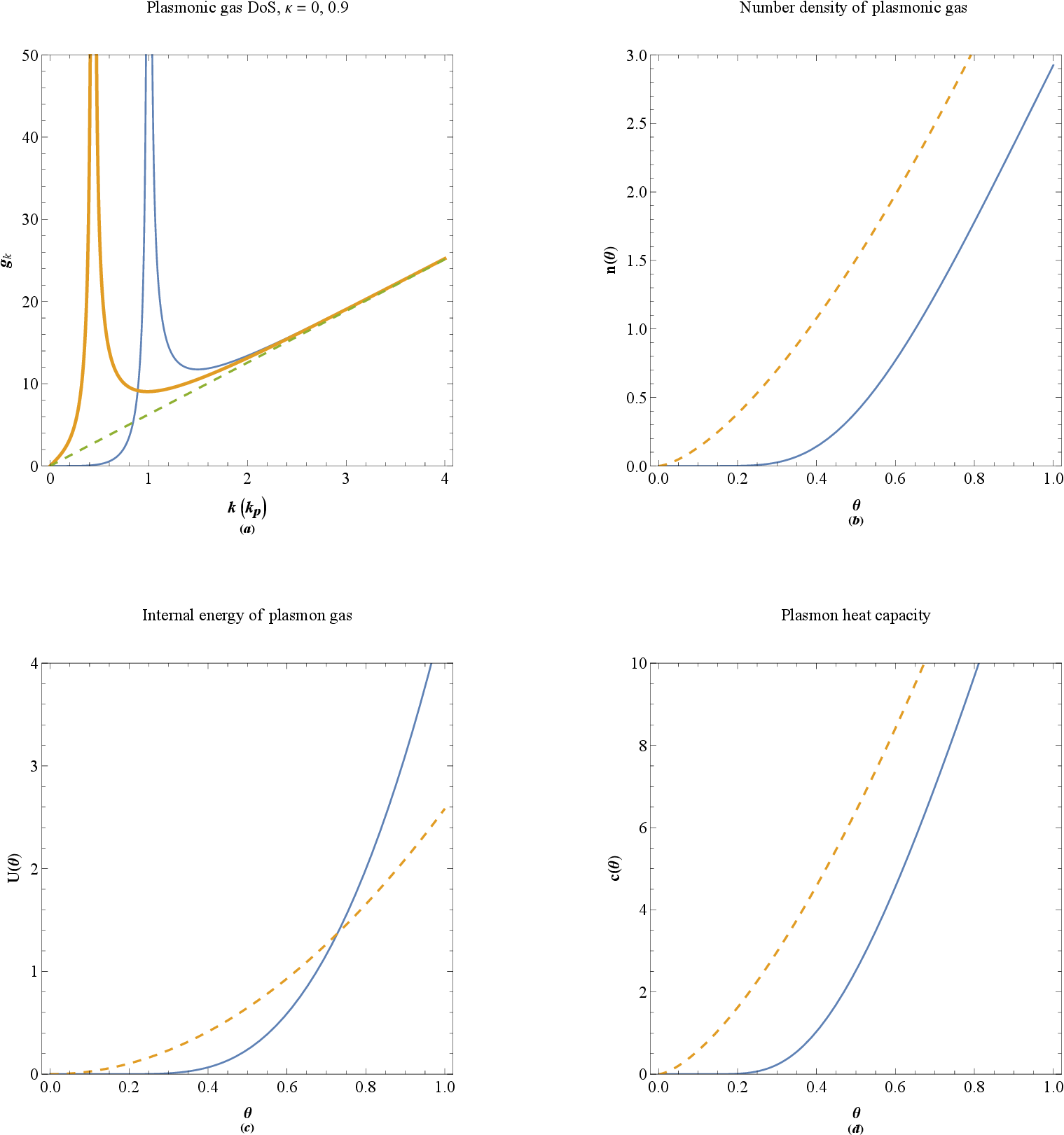}\caption{(a) Density of state of undamped plasmon (thin curve) and damped plasmon (thick curve) excitations. (b) The plasmon quasiparticle number density variation with normalized electron gas temperature. (c) The plasmonic gas internal energy variation with normalized electron gas temperature. (d) The plasmon excitation contribution to electron gas heat capacity as varied with normalized electron gas temperature.}
\end{figure}

Let us consider the dynamics of a quasineutral electron gas of finite temperature which can be described by the following effective Schr\"{o}dinger-Poisson system \cite{akbdual}
\begin{subequations}\label{sp}
\begin{align}
&i\hbar \frac{{\partial {\cal{N}}({\bf{r}},t)}}{{\partial t}} =  - \frac{{{\hbar ^2}}}{{2m}}\Delta {\cal{N}}({\bf{r}},t) - e\phi ({\bf{r}}){\cal{N}}({\bf{r}},t) + \mu {\cal{N}}({\bf{r}},t),\\
&\Delta \phi({\bf{r}}) = 4\pi e (|{\cal{N}}({\bf{r}},t)|^2-n_0),
\end{align}
\end{subequations}
where ${\cal{N}}({\bf{r}},t)=\psi({\bf{r}},t)\exp[iS({\bf{r}},t)/\hbar]$ denotes the many body statefunction characterizing the spatiotemporal evolution of the quatum electron fluid with the local electron number density $n({\bf{r}})=\psi({\bf{r}})\psi^*({\bf{r}})$ and $n_0$ being the uniform neutralizing positive background density. Furthermore, the electron fluid momentum is given by ${\bf{p}}({\bf{r}},t)=\nabla{S({\bf{r}},t)}$ and the equilibrium chemical potential $\mu_0$ is related to the local electron number density and the equilibrium temperature by an appropriate isothermal equation of state (EoS)
\begin{subequations}\label{np}
\begin{align}
&{n(\mu_0,T)} = \frac{{{2^{1/2}}m{^{3/2}}}}{{{\pi ^2}{\hbar ^3}}}  \int_{0}^{ + \infty } {\frac{{\sqrt{{\epsilon}} d{\epsilon}}}{{{e^{\beta ({\epsilon}-\mu_0)}} + 1}}},\\
&{P(\mu_0,T)} = \frac{{{2^{3/2}} m{^{3/2}}}}{{3{\pi ^2}{\hbar ^3}}}\int_0^{ + \infty } {\frac{{{{\epsilon}^{3/2}} d{\epsilon}}}{{{e^{\beta ({\epsilon} - {\mu_0})}} + 1}}.}
\end{align}
\end{subequations}
where $\beta=\mu_0/k_B T$ with $T$ being the electron temperature and $P$ the electron gas quantum statistical pressure. It is confirmed that the EoS chemical satisfies the thermodynamic identity $n\nabla\mu=\nabla P(n)$. In the EoS (\ref{np}) for the electron fluid the parameters $\mu_0$ and $T$ play the dominant role in description of state of the arbitrary degeneracy with he degeneracy parameter defined as $T_{F}/T$ in which $T_F$ is the Fermi temperature as related to the ratio $\mu_0/k_B T$. In the limit $T_F\ll T$ one has $\mu_0/k_B T\to -\infty$ which corresponds to the classical electron fluid whereas for $T_F> T$ we have $\mu_0/k_B T> 1$ which denotes the degenerate electron gas regime. On the other hand, for $T_F\gg T$ we have $\mu_0/k_B T\gg 1$ corresponding to the complete degeneracy limit in which $\mu_0\simeq E_F$ holds where $E_F=k_B T_F$ is the Fermi temperature, a parameter which is only dependent to the number-density of the electron fluid. Hence, the equation of state (\ref{sp}) covers a wide thermodynamic state of the electron gas.

Appropriately linearized system of (\ref{sp}) around the equilibrium state $\{\psi^0=1,\phi^0=0,\mu^0=\mu_0\}$ and consequent separation of space and time variables ${\cal{N}}({\bf{r}},t)\propto \exp[i({\bf{k}\cdot{\bf{r}}}-\epsilon t)]$ and ${\phi}({\bf{r}})\propto\exp[i({\bf{k}\cdot{\bf{r}}})]$ leads to coupled differential equation system
\begin{subequations}\label{pf}
\begin{align}
&{i}{\hbar }\frac{{\partial \Psi (t)}}{{\partial t}} =  E\Psi (t),\\
&\Delta \Psi ({\bf{r}}) + \Phi ({\bf{r}}) =  - 2E\Psi ({\bf{r}}),\\
&\Delta \Phi ({\bf{r}}) - \Psi ({\bf{r}}) = 0,
\end{align}
\end{subequations}
where ${\cal{N}}({\bf{r}},t)=\Psi({\bf{r}})\Psi(t)$ and $E=\epsilon-\sigma$ in which the energy $\epsilon$ is the normalized energy eigenvalues and $\sigma=\mu_0/E_p$. The energy eigenvalues $\epsilon$ and wavenumber $k$ are normalized, respectively to the plasmon energy $E_p=\sqrt{4\pi e^2 n_0/m}$ and $k_p=\sqrt{2m E_p}/\hbar$. Fourier analysis of the system (\ref{pf}) leads to the energy dispersion of collective excitations $E=k_w^2+k_e^2$ where $k_w$ and $k_e$ respectively denote the wave- and particle-like excitation wavenumberes. Note that $E=(\epsilon-\mu_0)$ is the normalized energy of the system as measured from the Fermi level. The dual wavenumber character of collective oscillations is an intrinsic feature of quantum electron fluid which has been extensively studied recently \cite{akbdual}. The wave-particle character of oscillations admit an interesting complementarity-like relation $k_w k_e=1$ in dimensionless form. The energy dispersion can be written in a more useful form of $E_k=k^2+1/k^2$ in which $k$ is the characteristic wavenumber of collective electron excitations. Therefore, the space and time variations are in respective scales of plasmon length $l_p=1/k_p$ and inverse plasmon frequency.

One may obtain thermodynamic quantities for quasiparticles of collective excitations called plasmons, sing the standard quantum statistical definitions. Assuming that there is a one-to-one correspondence between the single-electron and collective excitations energy levels \cite{akbquant}, as in Landau's quantum liquid states, the statistical equilibrium quantities for electron gas exactly translates into quasiparticle fluid quantities. Starting from the number of quasiparticle modes we have $N_k=4\pi k^3/3$ defining the density of states (DoS) $g_k=(dN_k/dk)/|dE_k/dk|$ with the quasiparticle occupation function of $f_k=1/[1+\exp(E_k/\theta)]$ in which $\theta=T/T_p$ being the normalized electron temperature as scaled to the plasmon temperature $T_p=E_p/k_B$. Therefore, the quasiparticle fluid of plasmonic excitations follow similar definitions as the electron gas, where the normalized number-density $n(\theta)$, internal energy $U(\theta)$ and heat capacity $c(\theta)$ are given as
\begin{subequations}\label{ther}
\begin{align}
&n(\theta ) = \int\limits_0^\infty  {\frac{{{g_k}{d_k}{E_k}dk}}{{1 + \exp ({E_k}/\theta )}}},\hspace{3mm}{g_k} = \frac{{2\pi {k^5}}}{{\left| {{k^4} - 1} \right|}},\\
&U(\theta ) = \int\limits_0^\infty  {\frac{{{g_k}{E_k}{d_k}{E_k}dk}}{{1 + \exp ({E_k}/\theta )}}},\hspace{3mm}{E_k} = {k^2} + \frac{1}{{{k^2}}},\\
&c(\theta ) = \frac{\partial }{{\partial \theta }}\int\limits_0^\infty  {\frac{{{g_k}{E_k}{d_k}{E_k}dk}}{{1 + \exp ({E_k}/\theta )}}},
\end{align}
\end{subequations}
The pseudoforce model (\ref{pf}) may be generalized to further include the pseudodamping element as follows \cite{akbdual}
\begin{subequations}\label{pf}
\begin{align}
&{i}{\hbar }\frac{{\partial \Psi (t)}}{{\partial t}} =  E\Psi (t),\\
&\Delta \Psi ({\bf{r}}) + 2\kappa \nabla \Psi ({\bf{r}}) + \Phi ({\bf{r}}) =  - 2E\Psi ({\bf{r}})\\
&\Delta \Phi ({\bf{r}}) + 2\kappa \nabla \Phi ({\bf{r}}) - \Psi ({\bf{r}}) = 0,
\end{align}
\end{subequations}
in which $\kappa$ is the normalized wavenumber of damping parameter. The later parameter may be used to model a variety of quantum phenomena such as charge screening and collective tunneling \cite{akbspill}. It can be shown that the generalized system admits the normalized dispersion $E(k,\kappa)=(k^2+\kappa^2)+1/(k^2+\kappa^2)$. The generalize quasiparticle DoS is then found to be $g(k,\kappa)=2\pi k(k^2+\kappa^2)/|(k^2+\kappa^2)-1|$. The latter may be compared to free electron density of state of $g_0(k)=2\pi k$ with the corresponding normalized energy dispersion $E_0(k)=k^2$.

Figure 1 shows variations of different plasmon parameters with electron number density. Figure 1(a) shows that the plasmon temperature increases exponential with increase of electron number density increasing from few Kelvins in semiconductors with low electron concentration up to $200000$K in metallic regime. Variation the normalized plasmon temperature with electron number density for different electron temperatures is depicted in Fig. 1(b). It is seen that this parameter sharply decreases with increase of electron concentration and has higher values for higher electron temperature. Figure 1(c) shows that plasmon temperature ranges from few tenth of $eV$ for semiconductors to a few $eV$ for metals and much higher for warm dense electron gas regime. The characteristic plasmon wavenumber variation in Fig. 1(d) reveals that this parameter also covers a wide range reaching up to $20$nm in metallic regime.

Figure 2 shows the collective quasiparticle energy dispersion and the effective electron mass in these excitations. Figure 2(a) depicts the undamped ($\kappa=0$) plasmoon dispersion. It is seen that the excitations are only stable for $E>2E_p$ with a minimum plasmon conduction occurring at characteristic plasmon wavenumber. Note that the fermi level coincides with $E=0$, i.e., $\epsilon=\mu$. For a given energy the double wavenumber excitation reaches a quantum beating state near the minimum of plasmon conduction $k_e\simeq k_w$. The dashed curve depicts the parabolic free electron dispersion. Figure 2(b) shows the energy dispersion of damped plasmons for $\kappa=0.5$. It is seen that the excitations become unstable above the critical quasiparticle energy $E_c=\kappa^2+1/\kappa^2$, hence, the collectrive excitations remain stable for $2<k<E_c$. Figure 2(c) depicts the effective electron mass ratio $m^*/m=[d^2E(k,\kappa)/dk^2]^{-1}$ for undamped plasmons $\kappa=0$. It is noted that the mass increases from $m=0$ at long wavelength limit to the electron rest mass for small wavelength (free electron) regime. Figure 2(d) reveals that the effective mass becomes negative for damped plasmon and diverges at a critical wavenumber $k=\sqrt {{{\left( {2{\kappa ^2} + \sqrt {1 + 4{\kappa ^4}} } \right)}^{1/3}} - {{\left( {2{\kappa ^2} + \sqrt {1 + 4{\kappa ^4}} } \right)}^{ - 1/3}} - {\kappa ^2}}$, i.e., $k\simeq 0.268$ for $\kappa=0.5$ in this plot.

Figure 3(a) depicts variation of the normalized plasmon DoS for undamped ($\kappa=0$) and damped with $\kappa=0.9$ with excitation wavenumber. It is seen that the DoS contains a Van-Hove-like singularity at $k=\sqrt{1-\kappa^2}$. Similar DoS singularity is well known to exist at the Fermi surface of crystalline materials. The dashed line shows the free electron gas DoS. Figure 3(b), on the other hand, shows the variation of the normalized quasiparticle (solid curve) and the free electron number density (dashed curve) for undamped excitations with respect to the normalized plasmon temperature. It is remarked that the quasiparticle number density is lower than the free electron value for all $\theta$ values. Figures 3(c) and 3(d) show the normalized internal energy and heat capacity compared for plasmon excitations (solid curve) and the free electron gas (dashed curve). It is seen that there are substantial differences between the two cases due to the difference in energy dispersion and the DoS.

\section{Lindhard Theory Beyond the Free-Electron Dispersion}

\begin{figure}[ptb]\label{Figure4}
\includegraphics[scale=0.6]{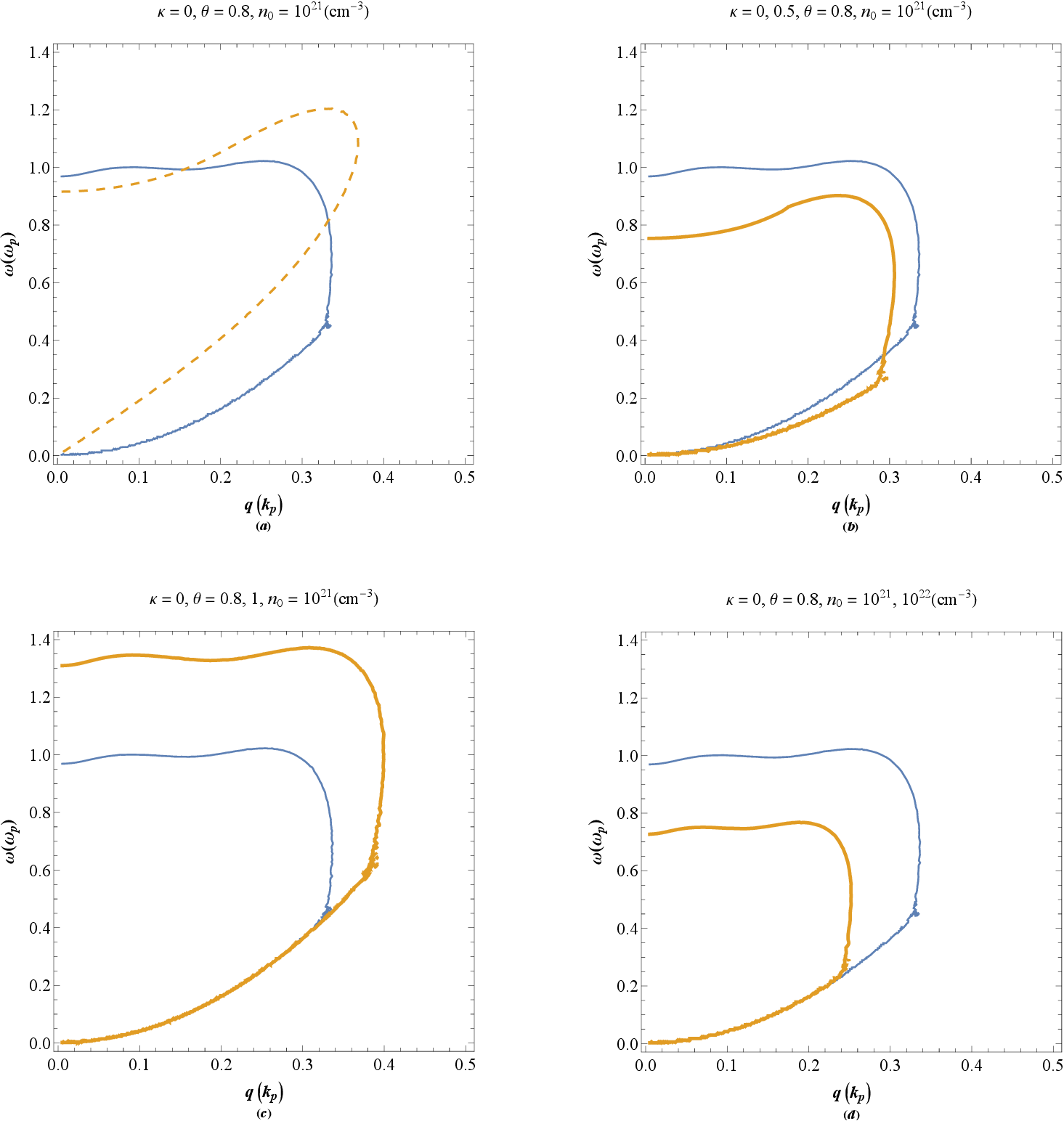}\caption{(a) Dispersion curve of finite temperature electron gas in conventional Lindhard model (dashed curve) and current model (solid curve). Effects of (b) the pseudodamping parameter (c) normalized electron gas temperature (d) the electron concentration on the electron dispersion in current model. The thicher curves correspond to higher parameter values above each panel.}
\end{figure}

In this section we would like to apply the findings of previous section to the Lindhard dielectric response \cite{lind} in order to get the plasmonic dielectric response of electron gas. To this end, we consider the equation (\ref{ld})
\begin{equation}\label{ldn}
\epsilon(q,\omega ) = 1 - \frac{{4\pi {e^2}}}{{{q^2}}}\left( {\frac{2}{V}} \right)\int {d^3{\bf{k}}} \frac{{{f_{{\bf{k}} + {\bf{q}}}} - {f_{\bf{k}}}}}{{\hbar \omega  + i\delta  - ({E_{{\bf{k}} + {\bf{q}}}} - E{_{\bf{k}}})}},
\end{equation}
where $V=(2\pi)^3$, $E_k=(k^2+\kappa^2)+1/(k^2+\kappa^2)$ and $f(k)=1/[1+\exp(E_k/\theta)]$. In the normalization scheme where frequency and wavenumber in units of plasmon quantities we have
\begin{equation}\label{ldnn}
\epsilon(q,\omega ) = 1 - \frac{{2{q_B}}}{{{q^2}{k_p}}}\int_{\beta\to 0^+} {k^2 dk}\int \sin(\theta)d\theta \frac{{{f_{{\bf{k}} + {\bf{q}}}} - {f_{\bf{k}}}}}{{\omega  + i\beta  - ({E_{{\bf{k}} + {\bf{q}}}} - E{_{\bf{k}}})}},
\end{equation}
where $q_B=1/r_B$ is the Bohr wavenumber with $r_B=\hbar^2/m e^2$ being the Bohr radius and $\beta=\delta/E_p$. A trivial change of variable leads to \cite{mih}
\begin{equation}\label{cv}
\epsilon(q,\omega ) = 1 - \frac{{2{q_B}}}{{{q^2}{k_p}}}\int_{\beta\to 0^+} {{f_{k}}{k^2}dk} \int {d\mu} \left( {\frac{1}{{\hbar \omega  + i\beta  -({E_{{\bf{k}} + {\bf{q}}}} - {E_{\bf{k}}})}} - \frac{1}{{\hbar \omega  + i\beta  +({E_{{\bf{k}} + {\bf{q}}}} - {E_{\bf{k}}})}}} \right),
\end{equation}
where $\mu=\cos(\theta)$, not to be confused with the chemical potential of the electron gas. Now using the following approximate expansion of energy dispersion, to the lowest orders in $\mu$, we have
\begin{equation}\label{exp}
{E_{k + q}} - {E_k} \simeq q.{\nabla _k}{E_k} + \frac{1}{2}{q^2}\nabla _k^2{E_k} +  \cdots  = 2q\mu k\eta (k,\kappa ) + {q^2}\gamma (k,\kappa ) +  \cdots,
\end{equation}
where $\eta(k,\kappa)=1-1/(k^2+\kappa^2)^2$ and $\gamma(k,\kappa)=1+(3k^2+\kappa^2)/(k^2+\kappa^2)^3$. Performing the inner integration and using the following identity
\begin{equation}\label{pv}
{\lim _{\beta  \to {0^ + }}}\frac{1}{{X + i\beta }} = p.v.\frac{1}{X} - i\pi \delta (X),
\end{equation}
where $p.v.$ indicates the principal value and $\delta$ denotes the Dirac delta function to separate the real and imaginary parts, we have
\begin{equation}\label{rd}
\epsilon{_r}(q,\omega ) = 1 - \frac{{2{q_B}}}{{{k_p}{q^3}}}\int\limits_0^\infty  {\frac{{{f_k}kdk}}{\eta }\left[ {\ln \left| {\frac{{{q^2}\gamma  - 2kq\eta  + \omega }}{{{q^2}\gamma  + 2kq\eta  + \omega }}} \right| + \ln \left| {\frac{{{q^2}\gamma  - 2kq\eta  - \omega }}{{{q^2}\gamma  + 2kq\eta  - \omega }}} \right|} \right]}.
\end{equation}
where $\epsilon_r$ is the real part of the dielectric function. Note that the limit $\eta=\gamma=1$ and $f_k=1/[\exp(k^2/\theta)+1]$ corresponds to free electron gas. And the imaginary part follows
\begin{subequations}\label{id}
\begin{align}
&\epsilon_i(q,\omega ) = \frac{{2{q_B}}}{{{k_p}{q^3}}}\int\limits_0^\infty  {\frac{{{f_k}kdk}}{\eta }\left[ {\Theta ( - kq\eta ,{q^2}\gamma  - 2kq\eta  - \omega , - {q^2}\gamma  - 2kq\eta  + \omega )} \right.}\\
&- \Theta ( - kq\eta , - {q^2}\gamma  - 2kq\eta  - \omega ,{q^2}\gamma  - 2kq\eta  + \omega )\\
&+ \Theta (kq\eta , - {q^2}\gamma  + 2kq\eta  - \omega ,{q^2}\gamma  + 2kq\eta  + \omega ) \\
& + \left. { \Theta (kq\eta ,{q^2}\gamma  + 2kq\eta  - \omega , - {q^2}\gamma  + 2kq\eta  + \omega )} \right].
\end{align}
\end{subequations}
where $\Theta(X)$ denotes the Heaviside theta function.

\begin{figure}[ptb]\label{Figure5}
\includegraphics[scale=0.6]{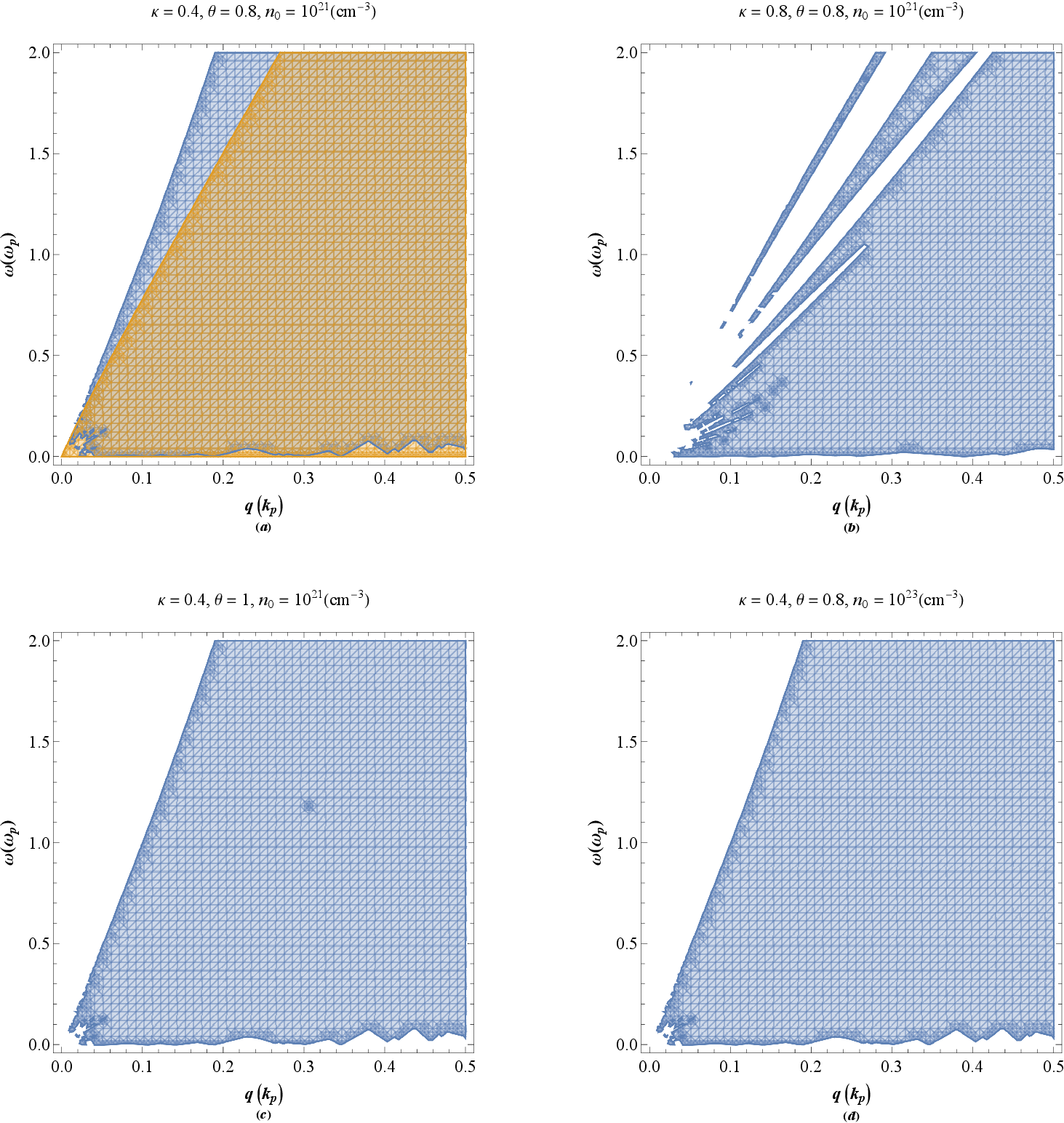}\caption{(a) The Landau damping region of finite temperature electron gas in conventional Lindhard model (small region) and current model (large region). Effects of (b) the pseudodamping parameter (c) normalized electron gas temperature (d) the electron concentration on the Landau damping region in current model.}
\end{figure}

\begin{figure}[ptb]\label{Figure6}
\includegraphics[scale=0.6]{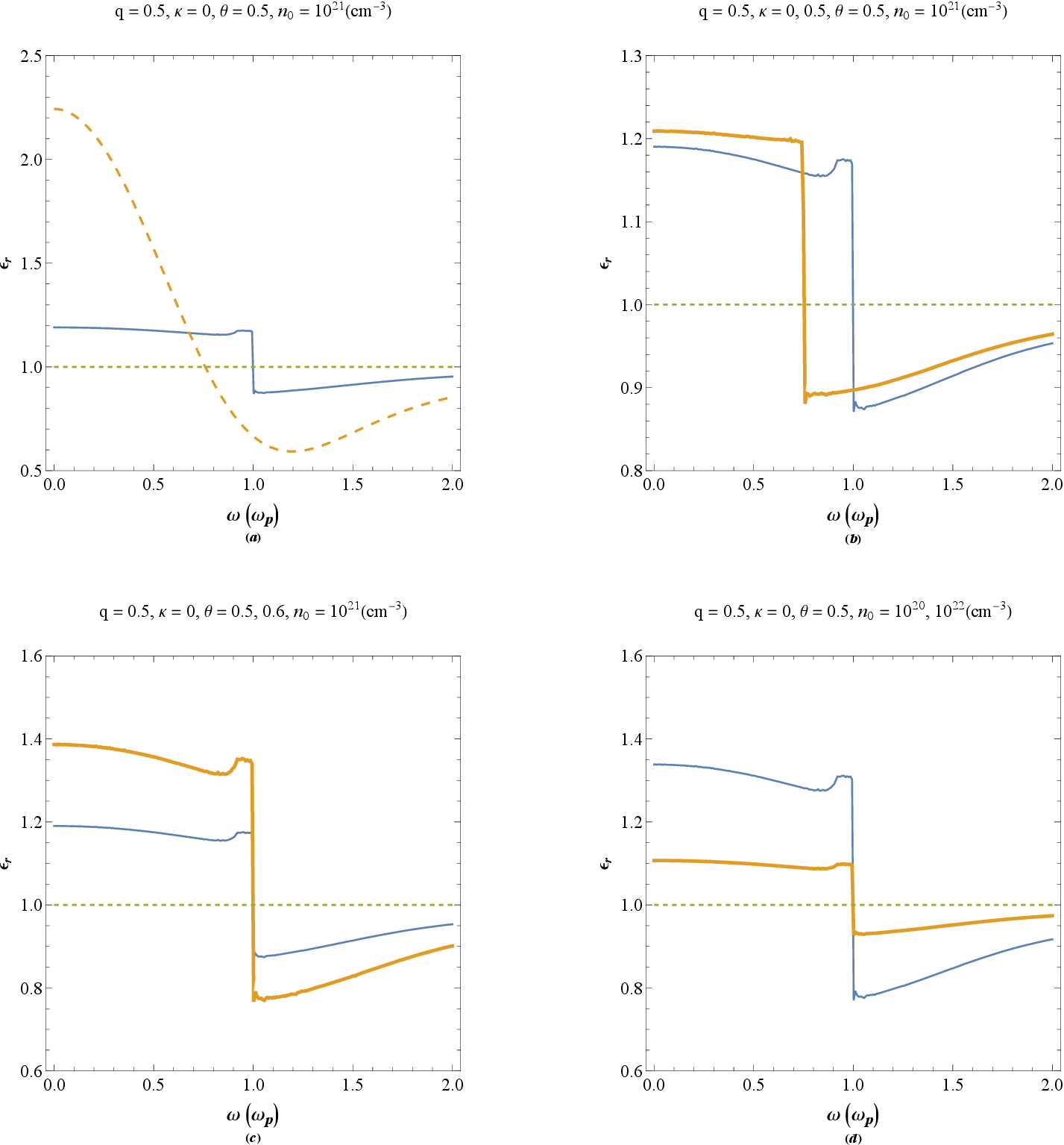}\caption{(a) The real part of dielectric function of finite temperature electron gas in conventional Lindhard model (dashed curve) and current model (solid curve). Effects of (b) the pseudodamping parameter (c) normalized electron gas temperature (d) the electron concentration on the real dielectric function in current model. The thicher curves correspond to higher parameter values above each panel.}
\end{figure}

Figure 4 depicts the plasmonic dispersion ($\epsilon_r=0$) curve of the Lindhard response theory for different electron gas parameters. In Fig. 4(a) the plasmonic dispersion (solid curve) is compared to that of the free electron dispersion (dashed curve) for undamped ($\kappa=0$). It is remarked that they while they agree at long wavelength limit, the plasmonic dispersion follows a wavy curve rather than parabolic shape for the free electron dispersion at intermediate wavelengths. However, the landau damping sets in for large wavenumbers where the two model diverge. Figure 4(b) shows the effect of pseudodamping parameter on plasmonic dispersion where the increase of this parameter significantly lower the response frequency for a given wavenumber. However, increase of the normalized electron gas temperature leads to significant increase in response frequency. The effect electron concentration on dispersion is shown in Fig. 4(d) indicating decrease of plasmonic response frequency for a Fermi gas with higher electron concentration. Note however that in this case the axes scales are also electron concentration dependent.

Figure 5 depicts the Landau damping \cite{krall} regions and its variation with different parameters shown above each plot. Figure 5(a) compares the plasmon model to that of free electron showing slightly larger region for plasmonic dispersion at higher response frequencies. It is remarked that for large values of pseudodamping parameter the Landau damping region is shattered. The Landau damping is due to poles of the denominator of integrand in dielectric response function. However due to moltipole structure in the case of plasmonic response the damping region for higher $\kappa$ values divides into multiple damping regions as shown in Fig 5(b). It is however concluded from Figs. 5(c) and 5(d) that the damping region is not significantly affected by electron temperature and concentration.

Figure 6 depicts the parametric variations of real dielectric function. Figure 5(a) indicates sharp resonance at plasmon frequency for current model (solid curve) as compared to the free electron Lindhard model (dashed curve). Figure 6(b) reveals that the resonance frequency is shifted to lower wavenumbers as the damping parameter increases. The increased value of electron gas temperature does not affect the resonance frequency, whereas, the amplitude of plasmonic dielectric variations is increased significantly. Also, the increase in value of electron gas concentration does not affect the resonance frequency, while, the amplitude of plasmonic dielectric variations is decreased significantly.

\begin{figure}[ptb]\label{Figure7}
\includegraphics[scale=0.6]{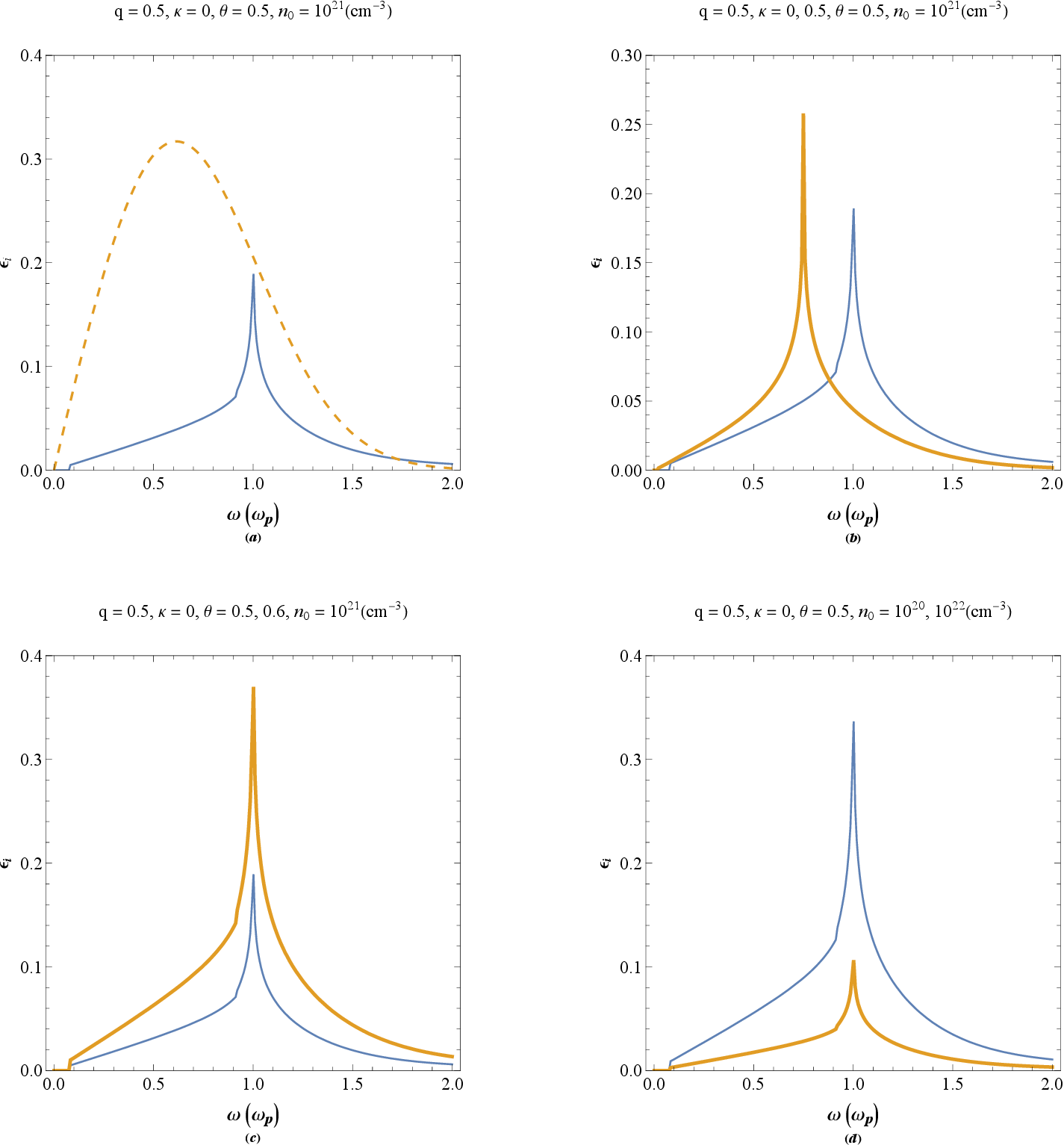}\caption{(a) The imaginary part of the dielectric function of finite temperature electron gas in conventional Lindhard model (dashed curve) and current model (solid curve). Effects of (b) the pseudodamping parameter (c) normalized electron gas temperature (d) the electron concentration on the imaginary dielectric function in current model. The thicher curves correspond to higher parameter values above each panel.}
\end{figure}

The corresponding imaginary part of plasmonic dielectric function to same parameter values in Fig. 6 is shown in Fig. 7. The plasmonic imaginary dielectric function (solid curve) sharply peaks at plasmon frequency which is completely different from that for free electron Lindhard model, as depicted in Fig. 7(a). The effects of pseudodamping parameter, electron temperature and concentration in the imaginary dielectric function, shown respectively in Figs. 7(b-d), follows the same behavior as the real part of Fig. 6. The real and imaginary parts of the dielectric function satisfy the relation $\epsilon(q,\omega)=\epsilon^*(q,-\omega)$.

\begin{figure}[ptb]\label{Figure8}
\includegraphics[scale=0.6]{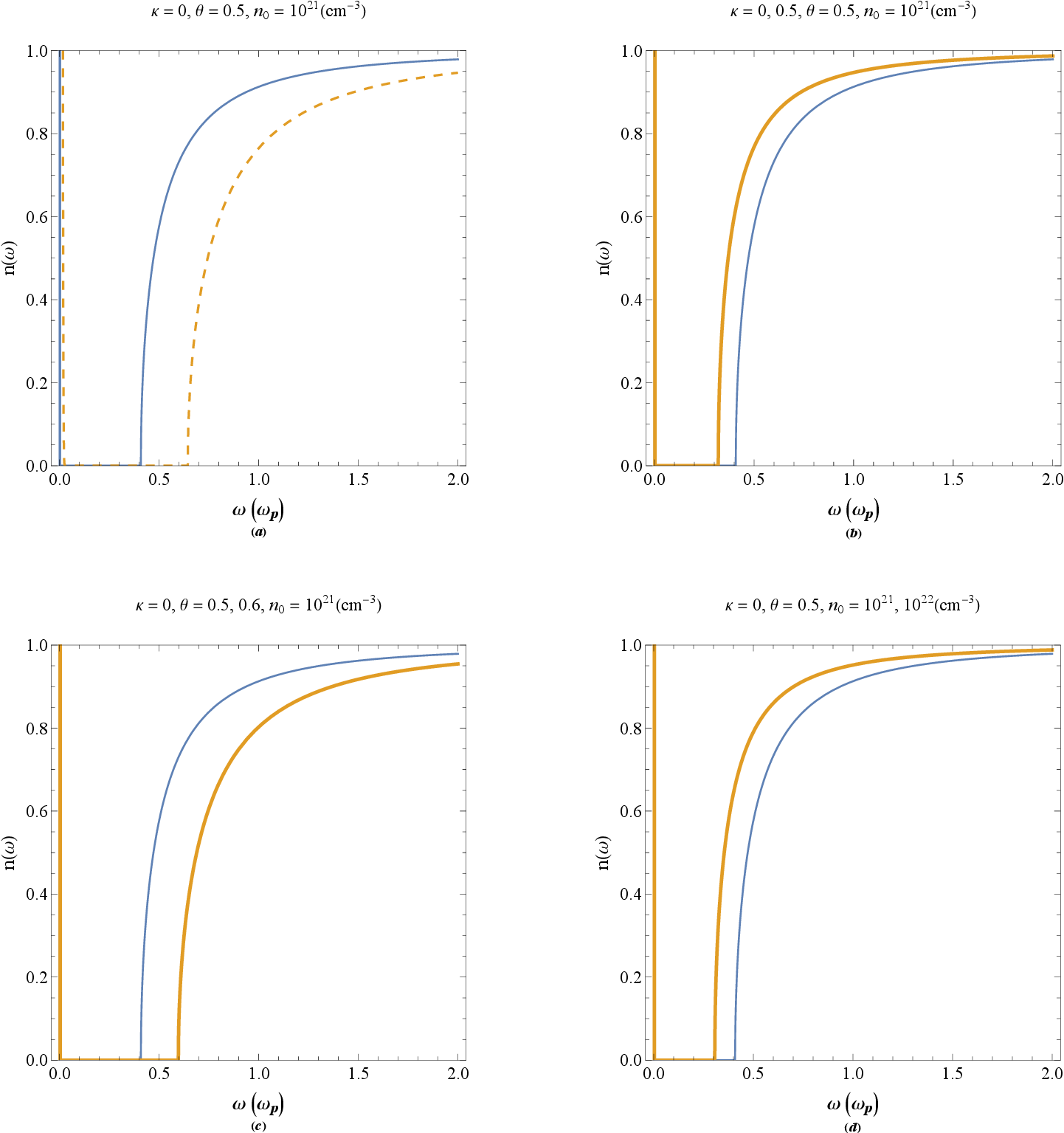}\caption{(a) The refraction index of finite temperature electron gas in conventional Lindhard model (dashed curve) and current model (solid curve). Effects of (b) the pseudodamping parameter (c) normalized electron gas temperature (d) the electron concentration on the index of refraction in current model. The thicher curves correspond to higher parameter values above each panel.}
\end{figure}

The real and imaginary parts of the dielectric function are also related via the Kramers-Kronig relations \cite{ichimaru3}
\begin{equation}\label{KK}
{\epsilon _r}(q,\omega ) = \frac{1}{\pi }p.v.\int_{ - \infty }^{ + \infty } {\frac{{{\epsilon _i}(q,\omega )}}{{\omega' - \omega }}} d\omega',\hspace{3mm}{\epsilon _i}(q,\omega ) =  - \frac{1}{\pi }p.v.\int_{ - \infty }^{ + \infty } {\frac{{{\epsilon_r}(q,\omega )}}{{\omega' - \omega }}} d\omega',
\end{equation}
where $p.v.$ refers to the Cauchy principal value of the integral. The dynamic conductivity of the electron gas is related to dielectric function by $\epsilon (\omega ) = 1 + 4\pi i\sigma (\omega )/\omega$ which  related to the conductivity in the classical Drude model via $\sigma(\omega)=\sigma_0/(1-i\omega\tau)$ where $\sigma_0=ne^2\tau/m$ and $n$ and $\tau$ refer to the electron number-density and collision time. Moreover, the complex index of refraction is given as $n(\omega)=\sqrt{\epsilon(\omega)}$. Therefore, the normal-angle optical reflectivity and absorption are given as
\begin{equation}\label{OPT}
R(\omega) = \frac{{{{\left[ {1 - {\Re}\sqrt {\epsilon (\omega)} } \right]}^2} + {\Im}\sqrt {\epsilon (\omega)} }}{{{{\left[ {1 + {\Re}\sqrt {\epsilon (\omega)} } \right]}^2} + {\Im}\sqrt {\epsilon (\omega)} }},\hspace{3mm}a(\omega) = \frac{{2\omega {\mathop{\Im}\nolimits} \sqrt{\epsilon(\omega)}}}{c},
\end{equation}
in which $\Re$ and $\Im$ respectively denote the real and imaginary parts. The parameter $a(\omega)$ is related to the electromagnetic wave penetration depth by $\xi(\omega)=2/a(\omega)$.

Figure 8 depicts the variations of real index of refraction. In figure 8(a) we see the significant difference between the predicted values in this (solid curve) and the free electron dielectric (dashed curve) models. Figures 8(b), 8(c) and 8(d) respectively depict the effects of pseudodamping parameter, the electron temperature and electron concentration on index of refraction of finite temperature electron gas in current model.

\begin{figure}[ptb]\label{Figure9}
\includegraphics[scale=0.6]{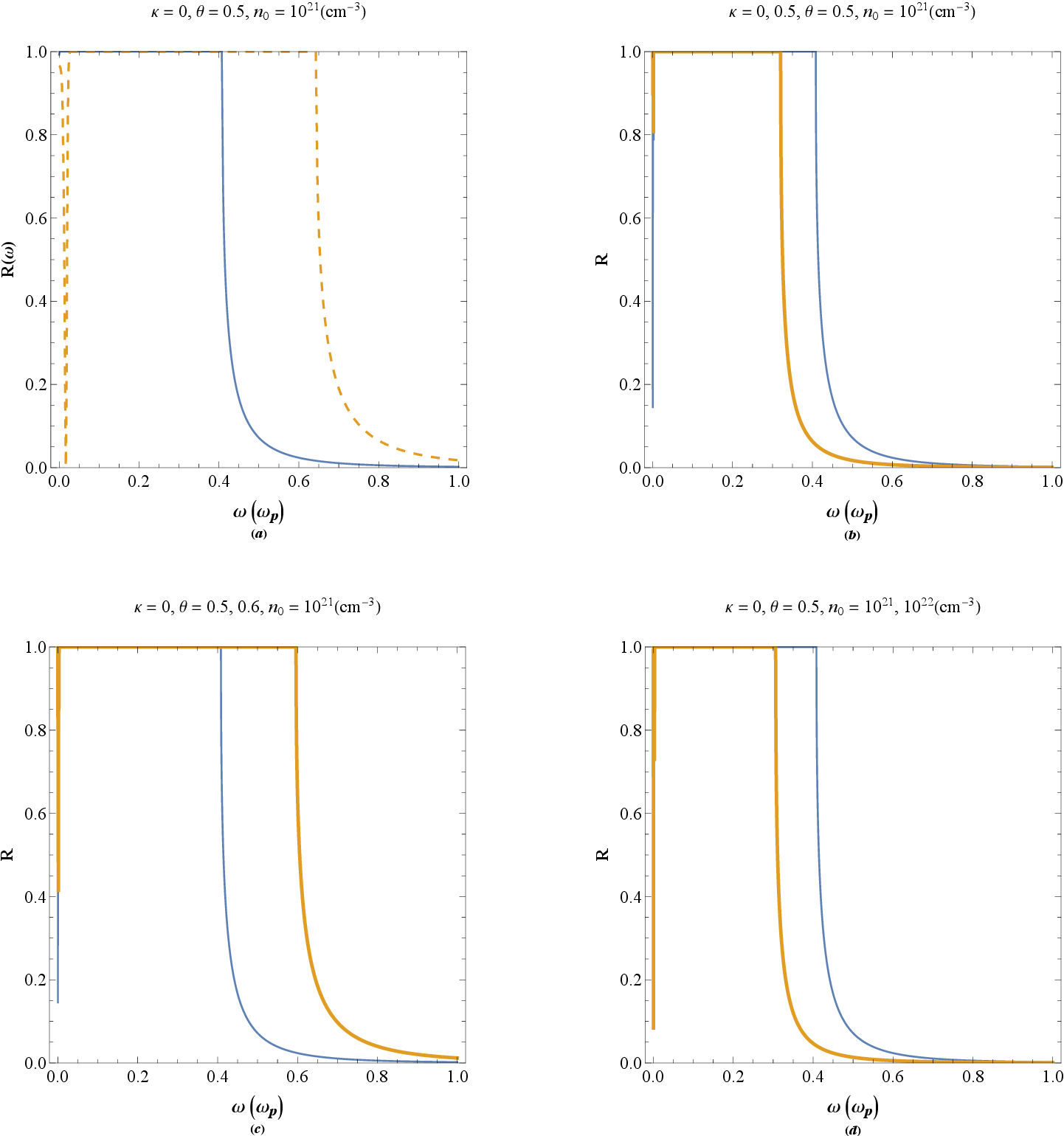}\caption{(a) The reflectivity of finite temperature electron gas in conventional Lindhard model (dashed curve) and current model (solid curve). Effects of (b) the pseudodamping parameter (c) normalized electron gas temperature (d) the electron concentration on the reflectivity in current model. The thicher curves correspond to higher parameter values above each panel.}
\end{figure}

The frequency dependent variation in the reflectivity of the electron gas is given in Fig. 9. The plasmon reflection edge is evident. Figure 9(a) shows that this edge occurs at significantly lower radiation frequency in our model as compared to the conventional Lindhard one. One also remarks that with increase of the pseudodamping parameter the reflection edge moves to lower frequencies, as shown in Fig. 9(b). Figure 9(c) and 9(d) indicates that the increase in the electron temperature causes the optical plasmon edge to significantly shift to higher frequencies and increase on electron concentration has the converse effect. The reflection edge is an important parameter in re-entry satellite communication blackout effects \cite{mos}.

\begin{figure}[ptb]\label{Figure10}
\includegraphics[scale=0.6]{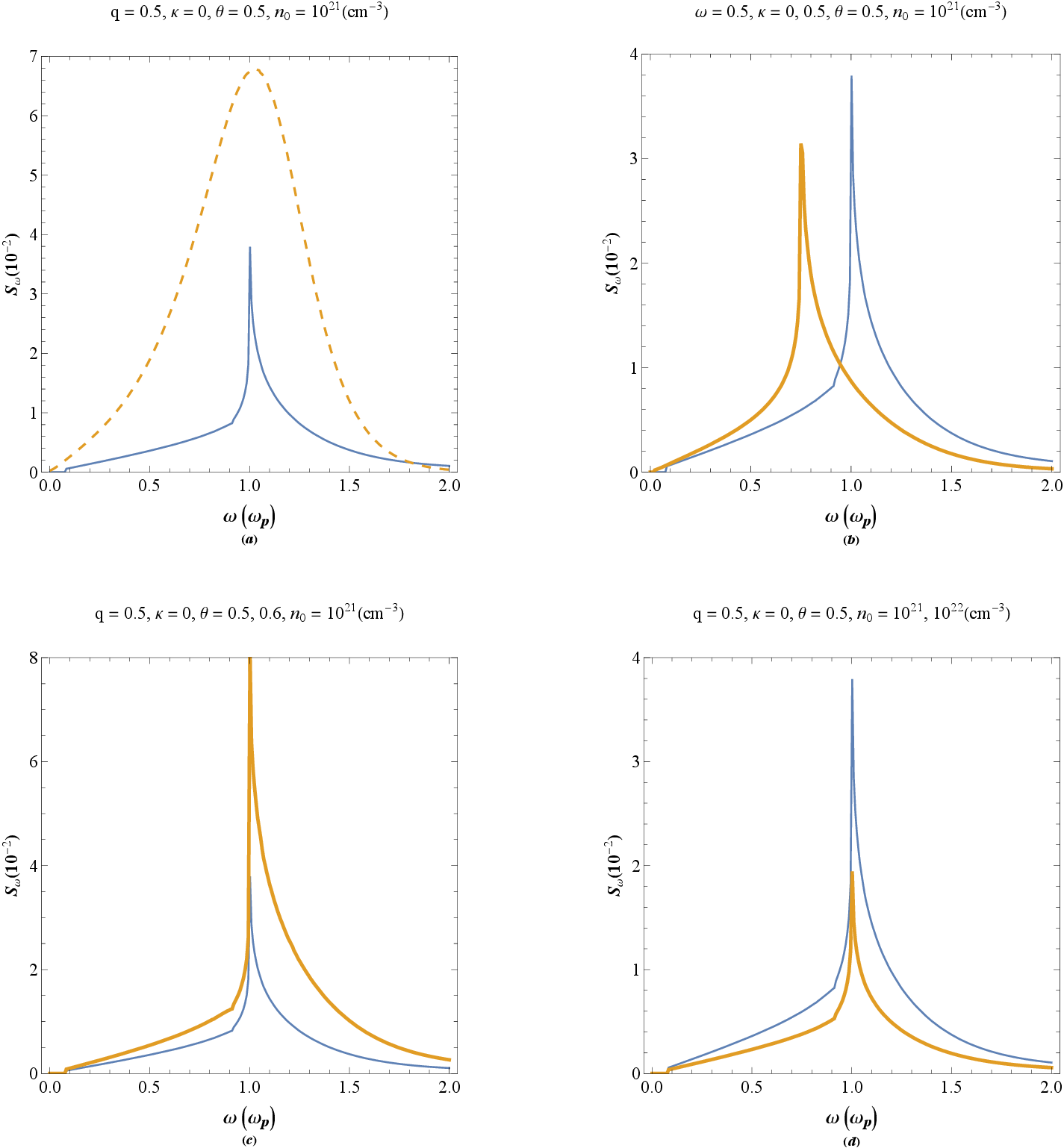}\caption{(a) The dynamic structure factor of finite temperature electron gas in conventional Lindhard model (dashed curve) and current model (solid curve). Effects of (b) the pseudodamping parameter (c) normalized electron gas temperature (d) the electron concentration on the dynamic structure factor in current model. The thicher curves correspond to higher parameter values above each panel.}
\end{figure}

The dynamic structure factor is related to the density-density response function via the fluctuation-dissipation theorem \cite{soylom,sturm}
\begin{equation}\label{FD}
S(q,\omega ) = \frac{{\hbar {\mathop{\Im}\nolimits} \left[ {\Pi (q,\omega )} \right]}}{{n\pi \left[ {\exp ( - \hbar \omega /{k_B}T)} - 1 \right]}},
\end{equation}
or in terms of the dielectric function, we have
\begin{equation}\label{SO}
S(q,\omega ) = \frac{{\hbar {q^2}}}{{4{\pi ^2}{e^2}n\left[ {1 - \exp ( - \hbar \omega /{k_B}T)} \right]}}{\mathop{\Im}\nolimits} \left[ { - \frac{1}{{\epsilon (q,\omega )}}} \right].
\end{equation}
Considering the symmetry properties of the dielectric function, the loss function can be written as
\begin{equation}\label{Sym}
\Im \left[ {\frac{-1}{{\epsilon (q,\omega)}}} \right] = \frac{i}{{2}}\left[ {\frac{1}{{\epsilon (q,\omega )}} - \frac{1}{{\epsilon ( - q, - \omega )}}} \right].
\end{equation}
Moreover, the dynamic structure factor satisfies the sum rule
\begin{equation}\label{Sum}
\frac{1}{{2\pi }}\int\limits_{ - \infty }^{ + \infty } {\omega S(q,\omega )} d\omega  = \frac{{\hbar {q^2}}}{{2m}}.
\end{equation}
The static structure factor $S(q)$, on the other hand, indicates the density correlations
\begin{equation}\label{SS}
S(q) = \int\limits_{ - \infty }^{ + \infty } {S(q,\omega )} d\omega.
\end{equation}
This parameter is related to the radial distribution function $g(r)$ as
\begin{equation}\label{Sg}
S(q) = 1 + \rho \int {{e^{ - i{\bf{q}} \cdot {\bf{r}}}}\left[ {g(r) - 1} \right]{d^3\bf{r}}}  = 1 + \frac{{4\pi \rho }}{q}\int {\sin (qr)\left[ {g(r) - 1} \right]rdr},
\end{equation}
where $\rho$ denotes the electron charge density. Figure 10(a) depicts the normalized dynamic structure factor of the finite temperature electron gas as compared in the two model. The structure factor peaks at the plasmon frequency for both while in the current model (solid curve) there is a resonant sharp maximum which is evidently due to the dual tone nature of the electron gas response and the intrinsic electrostatic wave-particle interactions. The later aspects agrees well with the surface plasmon resonance data \cite{ory}. It is remarked from Fig. 10(b) that this resonance frequency shifted to lower values by increase of the pseudodamping parameter. Moreover, Figs. 10(c) and 10(d) reveal that the increase in electron temperature/concentration increase/decrease the dynamic structure value leaving the resonance frequency unchanged.

\begin{figure}[ptb]\label{Figure11}
\includegraphics[scale=0.6]{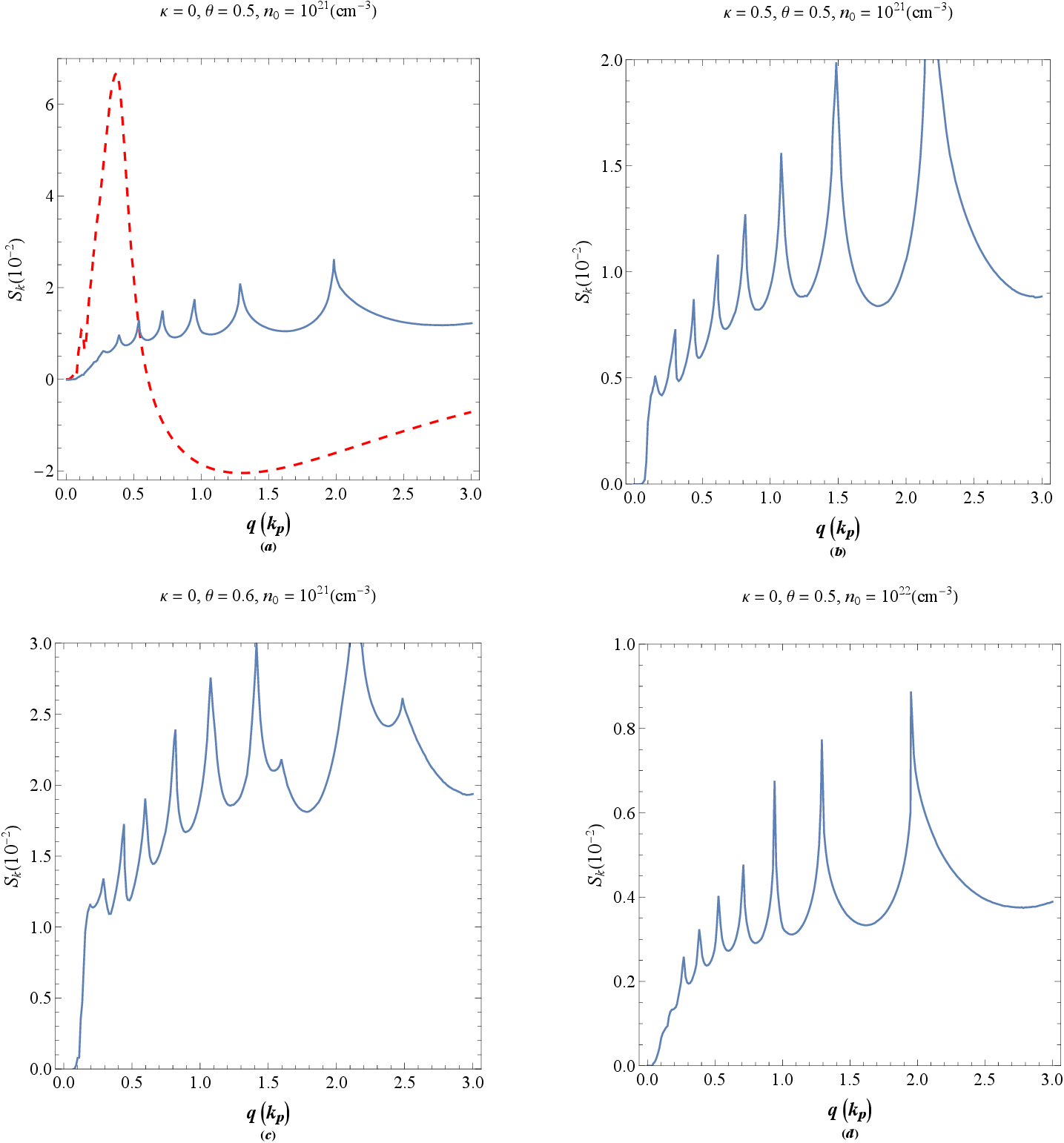}\caption{(a) The static structure factor of finite temperature electron gas in conventional Lindhard model (dashed curve) and current model (solid curve). Effects of (b) the pseudodamping parameter (c) normalized electron gas temperature (d) the electron concentration on static structure factor in current model.}
\end{figure}

Figure 11(a) shows the static structure factor of the electron gas in the two model. While the oscillations in the profile of current model (solid curve) indicates a strong density-density correlations, the profile corresponding to Lindhard theory with the single electron energy approximation (dashed curve) fails to correctly reflect such correlations. Recently, a hydrodynamic theory of dielectric density response has predicted a similar feature of the static structure factor for a degenerate electron gas in a wide density regime \cite{shmo}. it is remarked that increase in pseudodamping parameter leads to slight increase in frequency pattern of oscillation, as indicated by Fig. 11(b). Figure 11(c) shows the effect of electron temperature increase on correlation function indicating an interesting new dual tone correlation feature at small wavelength response. Furthermore, Fig. 11(d) indicates that the oscillation amplitude in correlation pattern is decreased by increase in electron concentration.

\begin{figure}[ptb]\label{Figure12}
\includegraphics[scale=0.6]{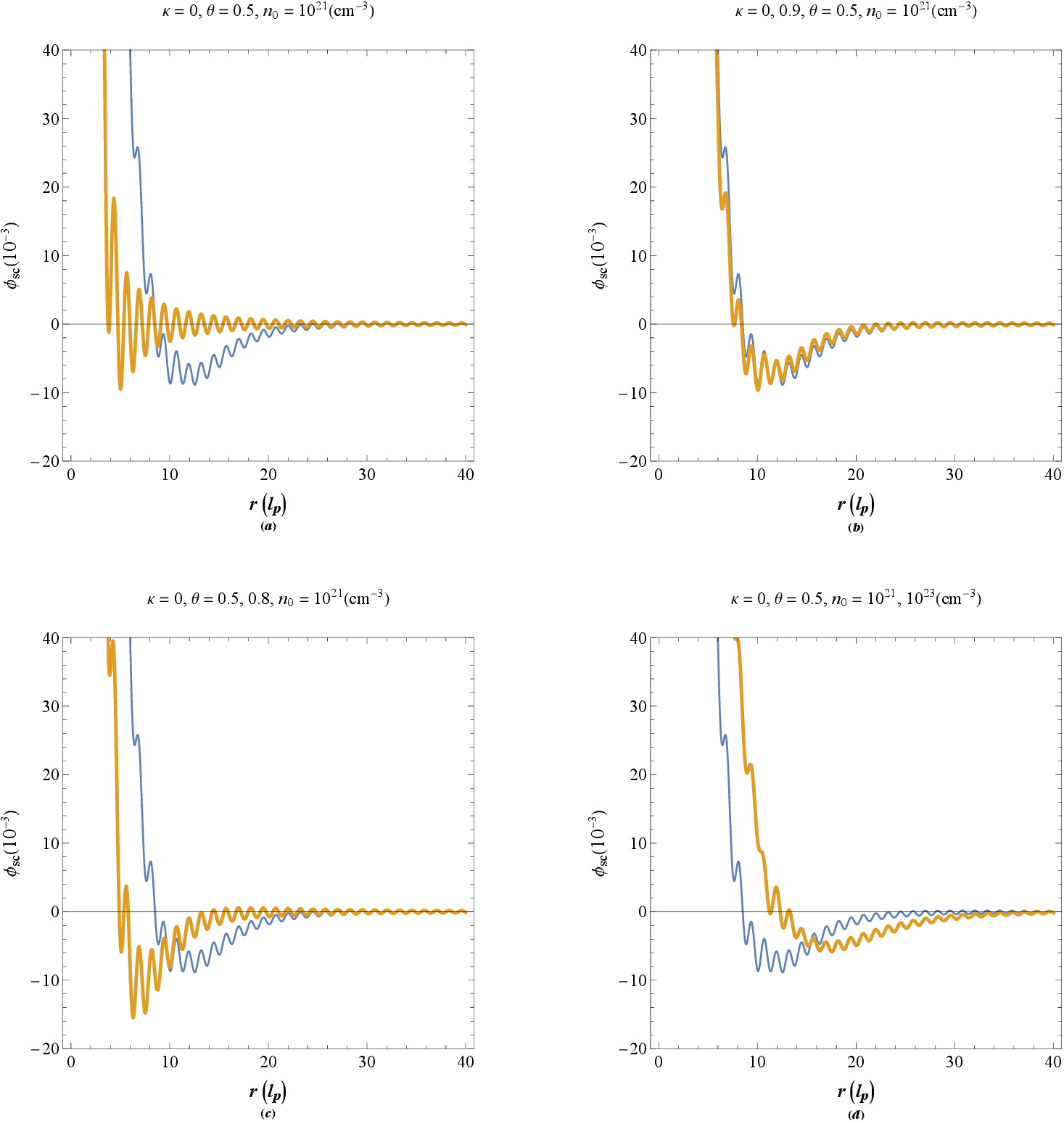}\caption{(a) The static screening potential of finite temperature electron gas in conventional Lindhard model (thick curve) and current model (thin curve). Effects of (b) the pseudodamping parameter (c) normalized electron gas temperature (d) the electron concentration on the static screening potential in current model. The thicher curves correspond to higher parameter values above each panel.}
\end{figure}

The static charge screening is another important property of the electron gas. in the classical model the screening potential follows the well-known Debye-Huckel theory \cite{chen}. However, for dense electron gas, where the inter-electron distances compare to de Broglie wavelength, the quantum effects dominate and charge screening follows the Fermi-Thomas theory \cite{ash} which incorporates the isothermal EoS for degenerate electron gas, as described previously. The Fermi-Thomas model however ignores important many-body effects such as the Friedel oscillations which are collective quantum electrostatic feature described via the Lindhard dielectric theory. Recently, attractive screening potential has been predicted around the impurity charge in the framework of quantum hydrodynamic theory \cite{seprl}. There has been intense debate \cite{bonitz1,sea1,bonitz2,sea2,bonitz3,akbarihd,sm,michta,moldabekov} over the past few years over the real nature of quantum screening which indicates that a kinetic correction factor must apply to the hydrodynamic Bohm factor in order to comply with the more precise Lindhard theory \cite{akbarihd}. The Lindhard screening potential does not resolve the attractive potential from the Friedel oscillations \cite{ash} with is due to interaction of electrostatic oscillation length-scale with the characteristic Fermi one. Current model however makes this distinction by making novel resolution between the long wavelength collective interactions and small wavelength single electron oscillation through the generalized energy dispersion relation and goes beyond the independent electron approximation for the dielectric response theory. The static charge screening is related to the dielectric response of the electron gas as
\begin{equation}\label{scr}
{\phi _{sc}}(r) = \frac{{Ze}}{{2{\pi ^2}}}\int {\frac{{\exp (i{\bf{q}} \cdot {\bf{r}}){d^3\bf{q}}}}{{{q^2}\epsilon_r(q,0)}}},
\end{equation}
where $Z$ the atomic number of screened ion. In Fig. 12(a) we compare the static screening potential for a singly ionized ($Z=1$) impurity in the two models. It is clearly evident that there are fundamental differences between current model prediction (solid curve) with that of the conventional Lindhard model (dashed curve). Current model predicts an oscillatory attractive screening potential (OASP) which clearly supersedes previous quantum hydrodynamic and Lindhard models by incorporation of dual length-scale response of electron gas treating the long and small wavelength perturbations in a separate wavenumber regimes. In the conventional Lindhard theory the small and large phase-speed static charge screening limits are approximated by appropriate expansion of the response or its inverse function \cite{akbarihd}. The exact solution of effective Schr\"{o}dinger-Poisson shows the similar feature around the impurity charge \cite{akbint}. Figure 12(b) shows that the pseudodamping parameter does not impact significantly the screening profile. Whereas, Fig.12(c) reveals that the screening potential valley moves close to the impurity charge and deepens as electron temperature rises. It is further remarked from Fig. 12(d) that increase of electron concentration shift the potential valley far away from the impurity charge. Existence of the attractive screening potential around impurity charge in electron gas may have a fundamental consequences for quantum ionic bound states in confinement fusion experiments \cite{krall}.

\begin{figure}[ptb]\label{Figure13}
\includegraphics[scale=0.6]{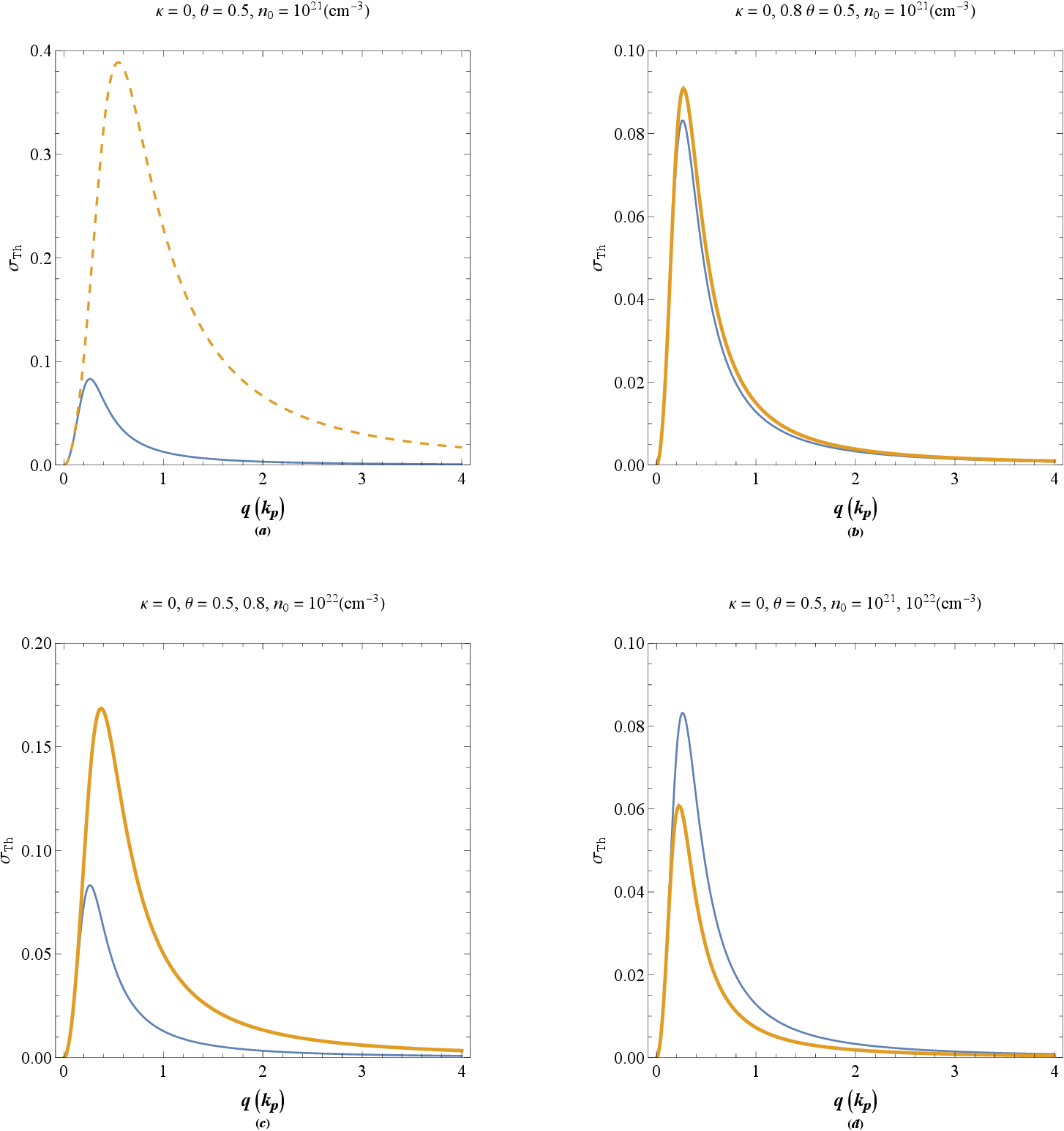}\caption{(a) The Thomson scattering cross section of finite temperature electron gas in conventional Lindhard model (dashed curve) and current model (solid curve). Effects of (b) the pseudodamping parameter (c) normalized electron gas temperature (d) the electron concentration on the Thomson scatterin cross section in current model. The thicher curves correspond to higher parameter values above each panel.}
\end{figure}

The Thomson scattering is an elastic scattering of electromagnetic waves off the electron gas. The dielectric response of electron gas is related to the angle resolved scattering cross section via \cite{jungnew}
\begin{equation}\label{ths}
{\sigma _{Th}} = \frac{1}{2}\int\limits_{ - 1}^{ + 1} {S({q_\mu},0)(1 - \xi )(1 + {\mu^2})d\mu},\hspace{3mm}S({q_\mu},0) = 2{q_{\mu}^2}\left[ {1 - \frac{1}{{\epsilon_r({q_\mu },0)}}} \right],
\end{equation}
where $\mu=\cos\theta$ and $q_{\mu}=q\sqrt{(1 - \mu)/2}$. Figure 13(a) compares the Thomson scattering cross section in the two different models. It is remarked that the cross section is greatly over estimated in Lindhard theory (dashed curve) with free electron energy approximation, as compared to current model (solid curve). The maximum scattering wavenumber is also slightly lower in current model. Figure 13(b) shows that the cross section slightly increases by increase in the pseudodamping parameter without significant shift in maximum scattering wavelength. It is also remarked that increase of electron temperature/concentration increase/decrease the Thomson scattering cross section, significantly, shifting maximum scattering wavenumber to higher/lower values.

\begin{figure}[ptb]\label{Figure14}
\includegraphics[scale=0.6]{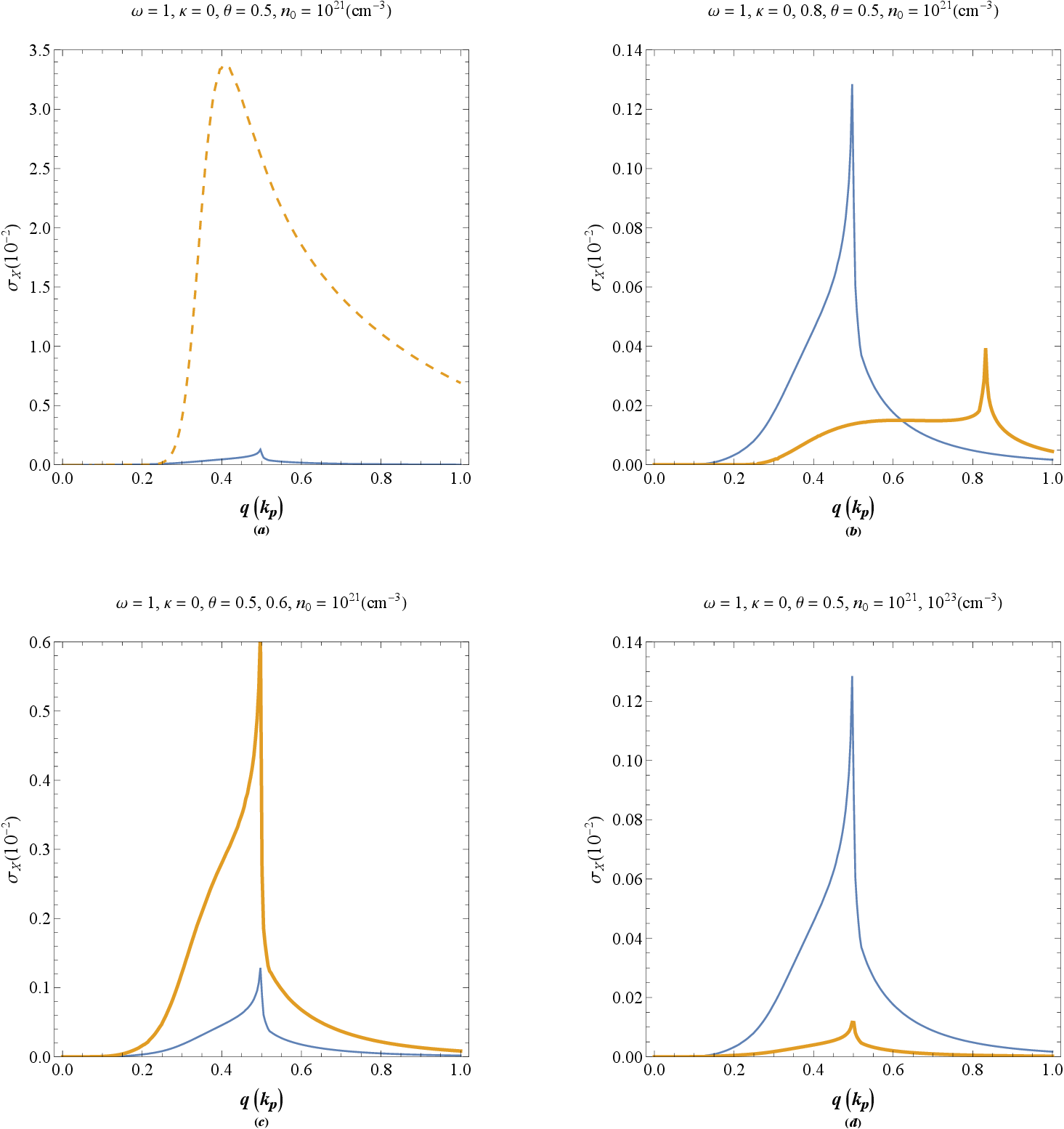}\caption{(a) The X-ray scatering cross section of finite temperature electron gas in conventional Lindhard model (dashed curve) and current model (solid curve). Effects of (b) the pseudodamping parameter (c) normalized electron gas temperature (d) the electron concentration on the X-ray scattering cross section in current model. The thicher curves correspond to higher parameter values above each panel.}
\end{figure}

The inelastic scattering of electromagnetic waves from the electron gas is related to the loss function as
\begin{equation}\label{SC}
\frac{{{d^2}{\sigma _R}}}{{d\Omega d\omega }} \propto  - \coth \left( {\frac{\hbar\omega }{{2k_B T}}} \right)\Im \left[ {\frac{1}{{\epsilon (q,\omega )}}} \right].
\end{equation}
The X-ray scattering is a inelastic scattering which is related to the both Thomson scattering cross section and the dynamic structure factor via a simple relation $\sigma_{X}=\sigma_X S(q,\omega)$. Variation X-ray scattering cross section is depicted in Fig 14. Figure 14(a) reveals a huge difference between the two model. The standard Lindhard theory (dashed curve) greatly overestimates this quantity. However, current model of X-ray scattering shows a sharp peak at half the plasmon wavenumber. The maximum X-ray scattering shifted to lower wavelengths by increase in pseudodamping parameter, as depicted in Fig. 14(b). Figures 14(c) and 14(d) indicate that the X-ray scattering cross section magnitude is significantly affected by changes in electron gas temperature and concentration.

\begin{figure}[ptb]\label{Figure15}
\includegraphics[scale=0.6]{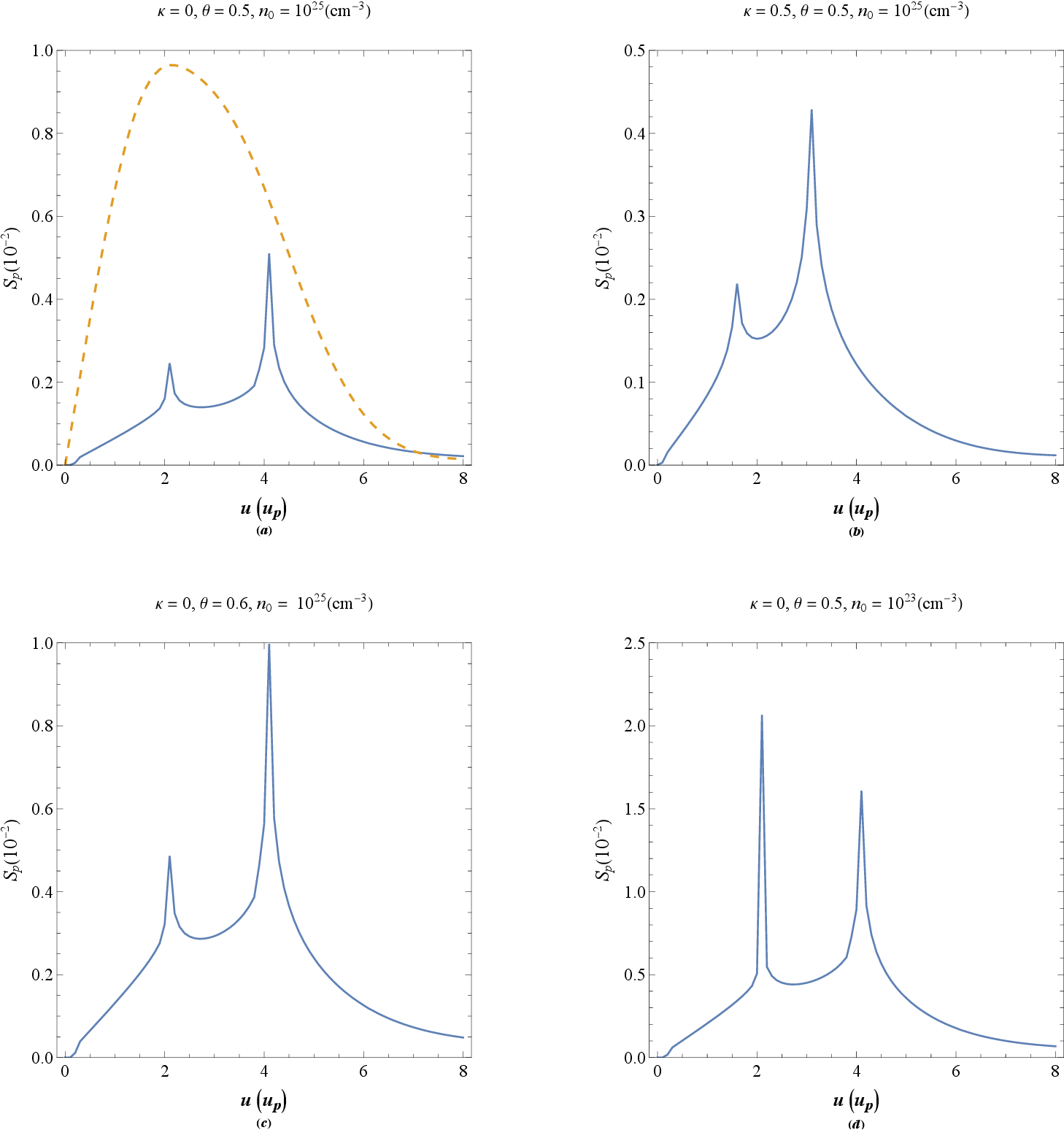}\caption{(a) The electronic stopping power of finite temperature electron gas in conventional Lindhard model (dashed curve) and current model (solid curve). Effects of (b) the pseudodamping parameter (c) normalized electron gas temperature (d) the electron concentration on the electronic stopping power in current model.}
\end{figure}

Finally, an important property of the electron gas is the power of stopping of energetic ion beam going through it. The stopping power has important application in inertial confinement fusion as new source of energy. The hydrodynamic theory of stopping power has been recently given for wide range of the electron number density and temperature \cite{shmo}. The basic calculation of this parameter is due to polarization and dielectric response of the plasma leading to the inelastic scattering effect. The electronic scattering rate is given as ${S_r}(q,\omega ) = {(4\pi{e^2}Z/{q^2})^2}(2\pi /{\hbar ^2}){S}(q,\omega)$. On the other hand, the rate of the energy loss is given by the following simple relation
\begin{equation}\label{sp1}
\frac{{dE}}{{dt}} = \int {\frac{{{S_r}(q,\omega )}}{{2\pi }}} \hbar \omega {\bf{dq}} =  - {\omega _p}{\left( {\frac{{Ze}}{\pi }} \right)^2}\int {\frac{{\omega \Im \left[ {{{1/\epsilon(q,\omega )}}} \right]{d^3\bf{q}}}}{{{q^2}\left[ {\exp \left( {\omega /\theta } \right) - 1} \right]}}}.
\end{equation}
The electronic contribution to the stopping power follows the classical relation $S_p=-dE/dl$ and is computed by the following double integral formula
\begin{equation}\label{sp2}
{S_p} =  - {\left( {\frac{{Z{\omega _p}}}{{v{v_p}}}} \right)^2}\int\limits_0^\infty  {\frac{{dq}}{q}} \int\limits_0^{qv} {\Im\left[ {{{\frac{1}{\epsilon(q,\omega )}}}} \right]\omega d\omega }.
\end{equation}
where $v$ is the ion speed and $v_p=\omega_p/k_p$ is the corresponding plasmon speed. Figure 15(a) shows the stopping power of the electron gas in the two model. It is seen that the stopping power in current model has two sharp ion beam speed stopping values whereas in conventional model there is a single maximum. The later may be due to the interaction of ion beam with both single electron oscillations as well as collective plasmon excitations. Figure 15(b) shows that increase in the pseudodamping parameter leads to an overal shift of stopping power to smaller speeds. Figures 15(c) reveals that increase in electron temperature does not modify maximum stopping speed but increase the magnitude stopping power. The decrease of electron concentration overall increase in stopping power as shown in Fig. 15(d).

\section{Conclusion}

In this research we provide a dual length-scale response theory of the finite temperature electron gas based on the generalized enrgy dispersion for electrons which incorporated their electrostatic interaction. The results of the current theory of dielectric response as compared to the Lindhard theory with free electron energy dispersion approximation reveals a fundamental difference between the two models. This was attributed to the dual wavenumber response of current model which incorporates both single electron oscillations as well as the collective plasmonic excitations. It was remarked that plasmonic frequency response of the electron gas is resonant and sharply peaks at the resonance frequency. We studied a large number of important physical quantities such as the dynamic and static structure factor, screening potential, dynamic optical properties, Thomson and X-ray scattering cross sections and electronic stopping power of the finite temperature electron gas in current and old models and investigated their variations in terms of various electron gas parameters. Current study leads to a better understanding of electron gas response to external electrodynamic perturbations for a wide range of nonrelativistic density-temperature regime. Current model also allows further extension to include electron-phonon interaction and electron exchange effects through the generalized electron energy dispersion in the framework of the Lindard dielectric response theory.

\section{Data Availability}

The data that support the findings of this study are available from the corresponding author upon reasonable request.


\section{References}

\begin{thebibliography}{}
\bibitem{lucio} L. C. Andreani, Physics Today \textbf{67}, 53(2014); doi.org/10.1063/PT.3.2386.
\bibitem{mark} P. A. Markovich, C.A. Ringhofer, and C. Schmeister, Semiconductor Equations (Springer, Berlin, 1990).
\bibitem{haug} H. Haug and S. W. Koch, "Quantum theory of the optical and electronic properties of semiconductors", World Scientific, 2004,
\bibitem{gardner} C. Gardner, SIAM, J. Appl. Math. \textbf{54} 409(1994).
\bibitem{man1} G. Manfredi, Phys. Plasmas \textbf{25}, 031701(2018); https://doi.org/10.1063/1.5026653
\bibitem{maier} S. A. Maier, Plasmonics: Fundamentals and Aplications, Springer Science Business Media LLC (2007).
\bibitem{yofee} A. D. Yofee, Adv. Phys., \textbf{42}, 173-262(1993), DOI: 10.1080/00018739300101484
\bibitem{chen} F. F. Chen, Introduction to Plasma Physics and Controlled Fusion, 2nd ed. (Plenum Press, New York, London, 1984).
\bibitem{krall} N. A. Krall and A. W. Trivelpeice, "Principles of Plasma Physics", (San francisco Press, San francisco 1986).
\bibitem{ichimaru1} S. Ichimaru, Rev. Mod. Phys. {\bf 54}, 1017 (1982).
\bibitem{ichimaru2} S. Ichimaru, H. Iyetomi, and S. Tanaka, Phys. Rep. {\bf 149}, 91 (1987).
\bibitem{ichimaru3} S. Ichimaru, {\sl Statistical Physics: Condensed Plasmas} (Addison Wesely, New York, 1994).
\bibitem{fetter} A. L. Fetter and J. D. Walecka, Quantum Theory of Many-Particle Systems,. McGraw-Hill 1971.
\bibitem{mahan} G. D. Mahan, Many-particle physics, 2nd edition, chapter 5 (Plenum press, New York,
1990).
\bibitem{pin} D. Pines and P. Nozieres, The Theory of Quantum Liquids (Addison-Wesley, 1968).
\bibitem{axel} D. Axel Becke, "Perspective: Fifty years of density-functional theory in chemical physics". J. Chem. Phys. \textbf{140} A301(2014).
\bibitem{fischer} Ch. F. Fischer, "General Hartree-Fock program". Comp. Phys. Comm. \textbf{43} 355–365(1987).
\bibitem{haas1} F. Haas, {\sl Quantum Plasmas: An Hydrodynamic Approach} (Springer, New York, 2011).
\bibitem{man2} G. Manfredi, “How to model quantum plasmas,” Fields Inst. Commun. {\bf 46}, 263–287 (2005); in Proceedings of the Workshop on Kinetic Theory (The
Fields Institute, Toronto, Canada 2004): http://arxiv.org/abs/quant--ph/0505004.
\bibitem{man3} G. Manfredi, P. A. Hervieux, and J. Hurst, Rev. Mod. Plasma Phys. \textbf{3} (2019); doi.org/10.1007/s41614-019-0034-0
\bibitem {kit} C. Kittel, Introduction to Solid State Physics, (John Wiely and Sons, New York, 1996), 7th ed.
\bibitem {ash} N. W. Ashcroft and N. D. Mermin, Solid State Physics (Saunders College Publishing, Orlando, 1976).
\bibitem {hu1} C. Hu, Modern Semiconductor Devices for Integrated Circuits (Prentice Hall, Upper Saddle River, New Jersey, 2010) 1st ed.
\bibitem {seeg} K. Seeger, Semiconductor Physics (Springer, Berlin, 2004) 9th ed.
\bibitem{ko} M. Koenig, A. Benuzzi-Mounaix, A. Ravasio, T. Vinci1, N. Ozaki, S. Lepape, D. Batani, G. Huser, T. Hall, D. Hicks,
A. MacKinnon, P. Patel, H. S. Park, T. Boehly, M. Borghesi, S. Kar and L. Romagnani, Plasma Phys. Control. Fusion {\bf 47}, B441 (2005).
\bibitem{chandra} S. Chandrasekhar, ”An Introduction to the Study of Stellar Structure”, Chicago, Ill.
(The University of Chicago press), (1939),  p.392.
\bibitem {madelung} E. Madelung, Z. Phys., 40 322(1926).
\bibitem {fermi} E. Fermi and E. Teller, Phys. Rev. {\bf 72}, 399 (1947).
\bibitem {hoyle} F. Hoyle and W. A. Fowler, Astrophys. J. \textbf{132}, 565(1960).
\bibitem {bohm} D. Bohm and D. Pines, Phys. Rev. \textbf{92} 609(1953).
\bibitem {bohm1} Bohm, D. Phys. Rev. \textbf{85}, 166–179 (1952).
\bibitem {bohm2} Bohm, D. Phys. Rev. \textbf{85}, 180–193 (1952).
\bibitem {pines} D. Pines, Phys. Rev. \textbf{92} 609(1953).
\bibitem {levine} P. Levine and O. V. Roos, Phys. Rev, \textbf{125} 207(1962).
\bibitem {klimontovich} Y. Klimontovich and V. P. Silin, in Plasma Physics, edited by J. E. Drummond (McGraw-Hill, New York, 1961).
\bibitem {se} P. K. Shukla and B. Eliasson, Phys. Rev. Lett. \textbf{99}, 096401(2007).
\bibitem {sten} L Stenflo Phys. Scr. \textbf{T50} 15(1994).
\bibitem {ses} P. K. Shukla, B. Eliasson, and L. Stenflo Phys. Rev. E \textbf{86}, 016403(2012).
\bibitem {brod1} G. Brodin and M. Marklund, New J. Phys. \textbf{9}, 277(2007).
\bibitem {mark1} M. Marklund and G. Brodin, Phys. Rev. Lett. \textbf{98}, 025001(2007).
\bibitem {man4} N. Crouseilles, P. A. Hervieux, and G. Manfredi, Phys. Rev. B {\bf 78}, 155412 (2008).
\bibitem {fhaas} F. Haas, G. Manfredi, P. K. Shukla, and P.-A. Hervieux, Phys. Rev. B, \textbf{80}, 073301 (2009).
\bibitem {scripta} B. Eliasson and P. K. Shukla, Phys. Scr. \textbf{78}, 025503 (2008).
\bibitem{stenf1} L. Stenflo, Phys. Scripta \textbf{14}, 320(1967).
\bibitem{stenf2} L. Stenflo, and N. L. Tsintsadze, Astrophys. Space Sci. \textbf{64}, 513(1979).
\bibitem{stenf3} L. Stenflo, Phys. Scripta \textbf{23}, 779(1981).
\bibitem{stenf4} L. Stenflo and P. K. Shukla, Phys. Plasmas \textbf{6}, 1382(1991).
\bibitem {bonitz} Zh. A. Moldabekov, M. Bonitz, and T. S. Ramazanov, Phys. Plasmas \textbf{25}, 031903 (2018); doi.org/10.1063/1.5003910
\bibitem {manfredi} G. Manfredi and F. Haas, Phys. Rev. B {\bf 64}, 075316 (2001);
\bibitem {hurst} J. Hurst, K. L. Simon, P. A. Hervieux, G. Manfredi and F. Haas, Phys. Rev. B \textbf{93}, 205402(2016).
\bibitem {lind} J. Lindhard, Kgl. Danske Videnskab. Selskab, Mat.-Fys. Medd. \textbf{28}, (1954).
\bibitem {stern} F. Stern, Phys. Rev. Lett. 18, 546 (1967).
\bibitem {sturm} K. Sturm, Z. Naturforsch. \textbf{48a}, 233-242(1993).
\bibitem {ionst} P. K. Shukla and M. Akbari-Moghanjoughi, Phys. Rev. E \textbf{87}, 043106(2013).
\bibitem{akbquant} M. Akbari-Moghanjoughi, Phys. Plasmas, \textbf{26}, 012104 (2019); doi.org/10.1063/1.5078740
\bibitem{akbnew1} M. Akbari-Moghanjoughi, Plasmonics (2022); doi.org/10.1007/s11468-022-01712-w
\bibitem{akbnew2} M. Akbari-Moghanjoughi, Sci. Rep. \textbf{11}, 21099 (2021).
\bibitem{akbheat} M. Akbari-Moghanjoughi, Phys. Plasmas, \textbf{26}, 072106 (2019); doi.org/10.1063/1.5097144
\bibitem{soylom} Jeno Solyom, "Fundamentals of the physics of solids: Vol. 3 Normal, Broken-Symmetry and Correlated Systems", Springer Heidelberg Dordrecht London New York (2010).
\bibitem{akbspill} M. Akbari-Moghanjoughi, Phys. Plasmas \textbf{29}, 082112 (2022); doi.org/10.1063/5.0102151
\bibitem{akbdual} M. Akbari-Moghanjoughi, Phys. Plasmas, \textbf{26}, 112102 (2019); doi.org/10.1063/1.5123621
\bibitem{mih} B. Mihaila, Lindhard function of a d-dimensional Fermi gas, arXiv:1111.5337v1 [cond-mat.quant-gas] (2011).
\bibitem{mos} A. Mousavi, A. Esfandiari-Kalejahi, and M. Akbari-Moghanjoughi, Phys. Plasmas \textbf{23}, 073511 (2016); doi.org/10.1063/1.4958816
\bibitem{ory} Schnitzer Ory, Giannini Vincenzo, Maier Stefan A. and Craster Richard V. Proc. R. Soc. A. \textbf{472}, 20160258 (2016): doi.org/10.1098/rspa.2016.0258
\bibitem{shmo} P. K. Shukla and M. Akbari-Moghanjoughi, Phys. Rev. E \textbf{87}, 043106 (2013); doi.org/10.1103/PhysRevE.87.043106
\bibitem{seprl} P. K. Shukla and B. Eliasson, Phys. Rev. Lett. {\bf 108}, 165007 (2012); {\bf 108}, 219902 (E) (2012); {\it ibid.} {\bf 109}, 019901 (E) (2012).
\bibitem{bonitz1} M. Bonitz, E. Pehlke, and T. Schoof, Phys. Rev. E \textbf{87}, 033105 (2013).
\bibitem{sea1} P. K. Shukla, B. Eliasson, and M. Akbari-Moghanjoughi, Phys. Rev. E \textbf{87}, 037101 (2013).
\bibitem{bonitz2} M. Bonitz, E. Pehlke, and T. Schoof, Phys. Rev. E \textbf{87}, 037102 (2013).
\bibitem{sea2} P. K. Shukla, B. Eliasson and M. Akbari-Moghanjoughi, Phys. Scr. \textbf{87} 018202 (2013).
\bibitem{bonitz3} M. Bonitz, E. Pehlke, and T. Schoof, Phys. Scr. \textbf{88}, 057001 (2013).
\bibitem{akbarihd} M. Akbari-Moghanjoughi, Phys. Plasmas \textbf{22}, 022103 (2015); {\it ibid.} {\bf 22}, 039904 (E) (2015).
\bibitem{sm} L. G. Stanton and M. S. Murillo, Phys. Rev. E \textbf{91}, 033104 (2015); {\em ibid.} \textbf{91}, 049901 (E) (2015).
\bibitem{michta} D. Michta, F. Graziani, and M. Bonitz, Contrib. Plasma Phys. \textbf{55}, 437 (2015).
\bibitem{moldabekov} Zh. Moldabekov, T. Schoof, P. Ludwig, M. Bonitz, and T. Ramazanov, Phys. Plasmas \textbf{22}, 102104 (2015).
\bibitem{akbint} M. Akbari-Moghanjoughi, Phys. Plasmas, \textbf{26}, 062105 (2019); doi.org/10.1063/1.5090366
\bibitem{jungnew} G. W. Lee and Y.-D. Jung, Phys. Plasmas. \textbf{20}, 062108 (2013).
\end{thebibliography}
\end{document}